\documentclass[aps,prb,showpacs,twocolumn,amsfonts,amsmath,amsmath,superscriptaddress,letterpaper,10pt,floatfix]{revtex4-1}

\usepackage{hyperref}
\usepackage{cleveref}
\usepackage{bm}
\usepackage{graphicx}
\usepackage{color}
\usepackage{footmisc}

\newcommand{\req}[1]{Eq.~(\ref{#1})}
\newcommand{\reqs}[1]{Eqs.~(\ref{#1})}
\newcommand{\rref}[1]{(\ref{#1})}
\newcommand{\ii}{\mathrm{i}}
\newcommand{\ee}{\mathrm{e}}
\newcommand{\uu}[1]{(u_{\alpha})}
\newcommand{\vv}[1]{(v_{\alpha})}
\newcommand{\avg}[1]{\left\langle #1 \right\rangle}

\begin{document}

\title{Suppression of odd-frequency  pairing by  phase-disorder  in a nanowire coupled to Majorana zero modes}
\author{Dushko Kuzmanovski}
\affiliation{Department of Physics and Astronomy, Uppsala University, Box 516, S-751 20 Uppsala, Sweden}
\affiliation{NORDITA, Stockholm University, Roslagstullsbacken 23, SE-106 91 Stockholm, Sweden}
\author{Annica M.~Black-Schaffer}
\affiliation{Department of Physics and Astronomy, Uppsala University, Box 516, S-751 20 Uppsala, Sweden}
\author{Jorge Cayao}
\affiliation{Department of Physics and Astronomy, Uppsala University, Box 516, S-751 20 Uppsala, Sweden}
\date{\today}

\begin{abstract}
Odd-frequency superconductivity is an exotic phase of matter in
which Cooper pairing between electrons is entirely dynamical in nature. 
Majorana zero modes exhibit pure odd-frequency superconducting correlations 
due to their specific properties. 
Thus, by tunnel-coupling an array of Majorana zero modes to a spin-polarized wire, it is in principle possible to engineer a bulk one-dimensional odd-frequency spinless $s$-wave superconductor. We here point out that each tunnel coupling element, being dependent on a large number of material-specific parameters, is generically complex with sample variability in both its magnitude and phase. Using this, we demonstrate that, upon averaging over phase-disorder, the induced superconducting, including odd-frequency, correlations in the spin-polarized wire are significantly suppressed.
We perform both a rigorous analytical evaluation of the disorder-averaged $T$-matrix in the wire, as well as numerical calculations based on a tight-binding model, and find that the anomalous, i.e. superconducting, part of the $T$-matrix is highly suppressed with phase disorder. We also demonstrate that this suppression is concurrent with the filling of the single-particle excitation gap by smearing the near-zero frequency peaks, due to formation of bound states that satisfy phase-matching conditions between spatially separated Majorana zero modes. Our results convey important constraints on the parameter control needed in practical realizations of Majorana zero mode structures and suggest that the achievement  of a bulk 1D odd-$\omega$ superconductivity from MZMs demand full control of the system parameters.

\end{abstract}
\maketitle

\section{Introduction}
As was originally demonstrated by Berezinskii,~\cite{bere74} the superconducting pair amplitude of fermions may also be odd under exchange of time coordinates, or, equivalently, frequency $\omega$, which extends the usual classification of superconducting (SC) pair symmetries at equal times.~\cite{PhysRevLett.66.1533,PhysRevB.45.13125} Part  of the importance of odd-frequency (odd-$\omega$) pairing is based on it allowing for  highly unconventional SC dynamical correlations, even suggested to be an instance of hidden order.~\cite{Balatsky2017} Examples of odd-frequency pairing  include long-range proximity effect in superconductor-ferromagnet junctions,~\cite{PhysRevLett.86.4096,Kadigrobov01} paramagnetic Meissner effect,~\cite{PhysRevB.64.132507,PhysRevX.5.041021} and Majorana zero modes (MZMs) in topological SCs.~\cite{PhysRevB.87.104513,cayao2019odd}

MZMs are their own antiparticles and  emerge isolated as end states in certain topological SCs.~\cite{kitaev} They currently hold particular relevance for  potential applications in topological quantum computation.~\cite{kitaev,Franz2015,RevModPhys.80.1083,Sarma:16} Remarkably, MZMs are also a clear example of states that carry pure odd-$\omega$ spin-polarized correlations. This is due to their intrinsic nature of being their own antiparticles that implies that their  particle-hole propagator is at the same time a particle-particle propagator, which for a zero-energy state automatically acquire an odd-$\omega$ dependence.~\cite{PhysRevB.87.104513,PhysRevB.92.121404,PhysRevB.95.174516,cayao2019odd} 

A promising experimentally feasible platform for topological SC, and therefore for MZMs, is based on combining nanowires (NWs) with Rashba spin-orbit coupling, a strong magnetic field, and proximity induced $s$-wave SC.~\cite{PhysRevB.81.125318,PhysRevLett.105.077001,PhysRevLett.105.177002,Mourik:S12} Other proposals for MZMs include magnetic atoms on spin-orbit coupled SCs,~\cite{PhysRevB.84.195442,Nadj-Perge602} and proximity-coupled SC on metallic edge states of topological insulators.~\cite{PhysRevB.79.161408,ncomms575} These physical realizations have motivated an enormous interest since the initial experiments.\cite{Aguadoreview17,LutchynReview08,magnatoms,zhangreview,tkachov19review} 

Due to the vast experimental progress on MZMs, a promising direction to realize odd-$\omega$ SC is using MZMs. In fact, it was recently proposed how to engineer a bulk one-dimensional (1D) odd-$\omega$ spinless $s$-wave SC by coupling an array of MZMs to a spin-polarized wire (SPW).~\cite{Lee16} The setup was also shown to exhibit a paramagnetic Meissner effect, whereby an applied magnetic flux is enhanced rather than screened by the induced supercurrent. Additionally, the induced pure odd-$\omega$ SC state was demonstrated to be robust against disorder in the coupling coefficients, which, nonetheless, were assumed to be real.
However, under more realistic conditions any couplings to MZMs can easily become complex.~\cite{PhysRevLett.104.056402,PhysRevLett.103.237001,PhysRevB.94.235446,PhysRevB.96.045440} In fact, in an array of MZMs, it is very unlikely to have equal or even real valued couplings due to the large amount of parameters possible to tune in topological SC wires.\cite{Aguadoreview17,LutchynReview08,magnatoms,zhangreview,tkachov19review} 
 This then implies that it is crucial to analyze the robustness of the induced odd-$\omega$ pairing also under complex and different couplings, and in particular disorder with respect to the complex phase of the coupling terms. 

In this work, we consider an array of MZMs coupled to a SPW through generalized complex couplings. In particular, we focus on the role of disorder in the phases of the complex couplings. We find that, apart from local odd-$\omega$ pairing, there generically also exist non-local odd- and even-$\omega$ correlations. All these amplitudes correspond to the equal-spin triplet (spin-polarized) symmetry as they arise as an induced effect due to MZMs. 
\begin{figure}[!ht]
\centering
\includegraphics[width=.47\textwidth]{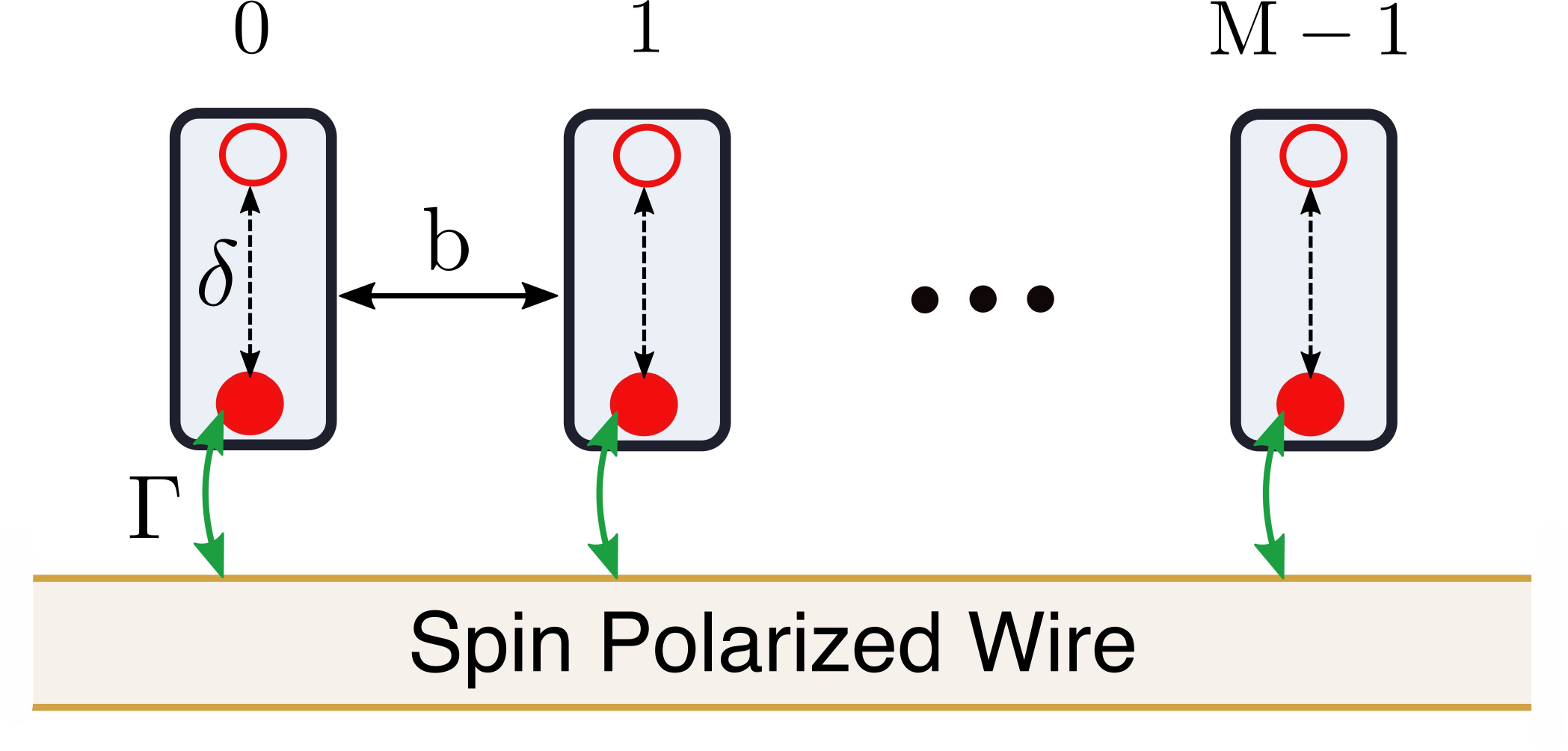} 
\caption{Sketch of the system where MZMs at one end (filled red circles) of 1D topological SC NWs  form an $M$-site long array, coupled through $\Gamma$ terms  to a SPW. MZMs at the other end (empty red circles) are not coupled to the SPW.  For finite length topological SCs, a finite energy splitting $\delta$ exists between two MZMs on each topological SC NW. The separation between each wire in the array is $b$.}
\label{Fig1}
\end{figure}
Most importantly, based on both analytical and numerical approaches, we demonstrate that even moderate phase disorder in the couplings drastically suppresses the induced SC in the SPW, in stark contrast to the effect of real disorder where even large variations of the couplings has been shown to only produce very small changes.\cite{Lee16} 
Furthermore, we show that the suppression of the induced pair correlations is accompanied by the filling of the energy gap in the density of states (DOS) due to formation of bound states in the SPW. As a consequence, our work shows that experimental realizations of a 1D odd-$\omega$ superconductor from an array of MZMs must have full control of the phases of the couplings.

The remainder of this article is organized as follows. In Sec.~\ref{sect1} we present the model, introduce the distribution of complex couplings, and discuss the Green's function approach for the pair correlations as well as a numerical tight-binding setup. In Sec.~\ref{sect3} we analytically calculate the pair amplitudes for one and two MZMs coupled to the SPW, and also perform a perturbative analysis for an array of MZMs. Complementary to this, in Sec.~\ref{sec:ResultsNumerics} we carry out a tight-binding numerical study in order to support our analytical findings. Finally, we summarize our work in Sec.~\ref{concl}. For completeness, we also provide additional supporting details in four Appendices.

\section{\label{sect1}Model and Method}
To create a 1D odd-$\omega$ superconductor we consider an SPW that is tunnel coupled to an array of MZMs. The MZMs are modeled as end modes of semiconducting NWs that are in a topologically non-trivial state, similar to Ref.~\onlinecite{PhysRevLett.104.056402}. Each NW carries one MZMs at each of its two ends, but only one of them is coupled to the SPW, as illustrated in Fig.~\ref{Fig1}.
The Hamiltonian $\mathcal{H}$ of the whole system is therefore composed of three parts $\mathcal{H} = \mathcal{H}_{w} + \mathcal{H}_{\Gamma} + \mathcal{H}_{\delta}$, where
\begin{equation}
\label{eq:WireHamil}
\mathcal{H}_{w}=\int \mathrm{d}x\, c_{x}^{\dagger} \, \hat{\xi}(x) \, c_{x},
\end{equation}
describes the SPW with $c^{\dagger}_{x}(c_{x})$ creating (destroying) a spin-polarized fermion in the  SPW and  $\hat{\xi}(x)=-(\hbar^{2}/2m)\partial^{2}_{x}-\mu_{w}$ is a ordinary parabolic dispersion relation, with $\mu_{w}$ being the chemical potential that determines the filling of the SPW.

The tunnel coupling of the SPW with MZMs at the topological NW end points is described by a tunneling Hamiltonian:\cite{PhysRevLett.104.056402}
\begin{equation}
\label{TunnelHamil}
\mathcal{H}_{\rm \Gamma}=\frac{i}{2}\sum_{n}\int dx \, \delta(x-n \, b) \, (\Gamma_{n}^{*} \gamma_{n}c_{x}+\Gamma_{n}\gamma_{n}c_{x}^{\dagger}),
\end{equation}
where $\Gamma_{n}$ represents the coupling strength between the MZM $\gamma_{n}$ and the fermionic mode in the SPW $c_{x}$ at position  $x_{n} = n \, b$, where $b$ is the separation between two neighboring MZMs. The MZM
$\gamma_{n} = f_{n} + f^{\dagger}_{n} = \gamma^{\dagger}_{n}$ is one of the two MZM formed from a zero-energy Bogoliubov excitation $f^{\dagger}_{n}$ of the topological NW.
For topological NWs of finite length  the two MZMs  split from zero-energy due to hybridization,\cite{PhysRevB.86.085408,PhysRevB.86.180503,PhysRevB.86.220506,PhysRevB.87.024515} which is commonly described by:\cite{kitaev}
\begin{equation}
\label{eq:HybridizationHam}
\mathcal{H}_{\delta} = \frac{1}{2} \sum_{n} \delta_{n} \, \left[f^{\dagger}_{n} \, f_{n} - f_{n} \, f^{\dagger}_{n}\right],
\end{equation}
where $\delta_{n}$ is the energy splitting in the $n$-th NW.\footnote{MZMs always appear in pairs, one at each end of a topological SC wire. If the length of this wire is less than twice the Majorana localization length, then the Majorana wavefunctions overlap, leading to a finite energy splitting $\delta$. Here, we couple only one these MZMs to the SPW but still consider finite $\delta$.}

We consider a general situation in which the coupling strengths $\Gamma_{n}$ are allowed to be all different from each other and also complex quantities, which describe  a  realistic situation for an array of MZMs coupled to a SPW.~\cite{PhysRevLett.104.056402,PhysRevLett.103.237001,PhysRevB.94.235446} Here, the SC phases in the topological NWs can easily be one of the sources for the complex $\Gamma$. Moreover, under realistic conditions, complex couplings $\Gamma$ might also appear just due to different parameters in the topological NWs. For example, recent studies in systems with spin-orbit coupling and magnetic field, have demonstrated  that the coupling between a MZM and normal systems (quantum dots) gives rise to a tunneling element that is both spin-dependent and complex.~\cite{PhysRevB.96.045440} 
Thus, since the couplings depend on the parameters of each NW and it is very hard to the imagine the NWs in an array all having exactly the same parameters, the couplings will generically be complex and vary randomly. We are therefore interested in analyzing the robustness of the induced SC correlations into the SPW against disorder in the couplings, and particularly phase disorder, not previously considered. For real and equal $\Gamma_{n}$, Eq.\,(\ref{TunnelHamil}) reduces to the model used in Ref.~\onlinecite{Lee16}, where real disorder was also studied.

More specifically, we model the complex couplings $\Gamma_{n} = \vert \Gamma_{n} \vert \, \ee^{\ii \, \theta_{n}}$ as random variables with the logarithm of the magnitude $\ln(\vert \Gamma_{n} \vert/\Gamma_{0})$ being uniformly distributed and the scaled phase $\theta_{n}/\pi$ being independently distributed according to a weighted uniform distribution around $0$ with weight $(1 + s)/2$ and around $1$ and $-1$ with weight $(1 - s)/4$, with in total the width $D_{\phi}$ of the $2\pi$ phase interval covered. We thus characterize phase disorder by $D_{\phi}$, where $D_{\phi}=1$ means full phase disorder. Physically, a finite value of $D_{\phi}$ indicates the lack of control over the phases in the complex couplings, an issue impossible to avoid experimentally.
We also introduce a sign bias, i.e.~weight of positive values minus weight of negative values, equal to $s, \ -1 \le s \le 1$. 
There is nothing fundamental in this choice of distribution, and, in fact, other distributions that model phase disorder would produce similar conclusions. However, this choice has the advantage of allowing the moments $\langle \left(\vert \Gamma \vert^{2} - \langle \vert \Gamma \vert^{2} \rangle \right)^{q} \rangle$, and $\langle \Gamma^{q} \rangle$ to be expressed analytically, exhibiting also a high suppression with increase in the power $q$ and width of the phase distribution $D_{\phi}$. In particular, the expectation value:
\begin{eqnarray}
& \avg{\ee^{\ii \, q \, \theta}} = \frac{1 - s}{2} \, \left\lbrace \frac{2}{D_{\phi}} \, \mathrm{sinc}(q) - \right. \nonumber \\
& \left. - \frac{2 - D_{\phi}}{D_{\phi}} \, \mathrm{sinc} \left[q \, \left(1 - \frac{D_{\phi}}{2}\right)\right] \right\rbrace + \frac{1 + s}{2} \, \mathrm{sinc}\left(\frac{q \, D_{\phi}}{2}\right), \label{eq:ExpAvg}
\end{eqnarray}
where $\mathrm{sinc}(x) = \frac{\sin(\pi \, x)}{\pi \, x}$. Another advantage of this distribution is that the case of real couplings strengths is achieved as a continuous limit by simply taking $D_{\phi} \rightarrow 0$. In that case, \req{eq:ExpAvg} reduces to $\left[(1 + s) + (1 - s) \, \cos(\pi \, q)\right]/2$, as expected for a distribution of real numbers with a sign bias $s$. 
More details on the parameters of the distribution are given in Appendix~\ref{sec:ModelPDF}.

\subsection{\label{DysonModel}Dyson's equation for the SPW}
The focus of this work is on the pair correlations induced in the SPW, and thus it is convenient to work with a Nambu-Gor'kov Green's function in imaginary time:
\begin{equation}
\label{eq:GreenFuncDef}
\hat{G}(x, \tau; x', \tau') = -\left\langle T_{\tau} \psi(x, \tau) \otimes \bar{\psi}(x', \tau') \right\rangle,
\end{equation}
where $\psi(x) = \left( c_{x}, c^{\dagger}_{x} \right)^{\top}$, and the imaginary-time Heisenberg operators are $\psi(x, \tau) = \ee^{\tau \, \mathcal{H}} \, \psi(x) \, \ee^{-\tau \, \mathcal{H}}$ and $\bar{\psi}(x', \tau') = \ee^{\tau' \, \mathcal{H}} \, \psi^{\dagger}(x') \, \ee^{-\tau' \, \mathcal{H}}$. 
With only the relative time $\tau - \tau'$ being relevant we perform a Fourier series expansion with respect to the relative time into Matsubara frequencies $\omega_{m} = (2 n + 1) \, \pi \, T$. 
In this way, the Green's function is a $2 \times 2$ matrix in particle-hole space:
\begin{eqnarray}
\label{eq:GreenFuncMat}
& \hat{G}(x, x'; \ii \, \omega_{m}) \nonumber \\
& = \left(\begin{array}{cc}
G_{ee}(x, x'; \ii \, \omega_{m}) & G_{eh}(x, x'; \ii \, \omega_{m}) \\
G_{he}(x, x'; \ii \, \omega_{m}) & G_{hh}(x, x'; \ii \, \omega_{m})
\end{array}\right).
\end{eqnarray}

The equation of motion for $\hat{G}(x, x'; \ii \, \omega_{m})$ requires the evaluation of the commutator of $\mathcal{H} = \mathcal{H}_{w} + \mathcal{H}_{\Gamma} + \mathcal{H}_{\delta}$, given by \reqs{eq:WireHamil}-\rref{eq:HybridizationHam}, with $c_{x}$ (and $c^{\dagger}_{x}$). When this is evaluated, terms proportional to $c_{x}$ (or $c^{\dagger}_{x}$) and $\gamma_{n} = f_{n} + f^{\dagger}_{n}$ are obtained. Thus, the equation of motion for $G$  contains the correlators $\langle\langle f_{n} \vert c^{\dagger}_{x}\rangle\rangle_{\omega}$, $\langle\langle f_{n} \vert c_{x}\rangle\rangle_{\omega}$, $\langle\langle f^{\dagger}_{n} \vert c^{\dagger}_{x}\rangle\rangle_{\omega}$, and $\langle\langle f^{\dagger}_{n} \vert c_{x}\rangle\rangle_{\omega}$. In order to eliminate these cross-correlators, another set of commutators, of $\mathcal{H}$ with $f_{n}$ (or $f^{\dagger}_{n}$), is used, which generates terms proportional to $f_{n}$ (or $f^{\dagger}_{n}$) and $\frac{\ii}{2} \, \delta(x - n \, b) \, \left[\Gamma_{n} \, c^{\dagger}_{x} + \Gamma^{\ast}_{n} \, c_{x} \right]$. In this way eliminating the cross-correlators, while using the Green's function for isolated NW modes:
\begin{eqnarray*}
& [\hat{G}_{\delta}]_{m n}(\ii \, \omega_{m}) = \langle\langle \left( \begin{array}{c}
f_{m} \\ f^{\dagger}_{m}\end{array} \right) \otimes \left( \begin{array}{cc}f_{n} & f^{\dagger}_{n}\end{array} \right)^{\top} \rangle\rangle_{\omega} \\
& = \delta_{m n} \, \left(\begin{array}{cc}
\frac{1}{\ii \, \omega_{m} - \delta_{n}} & 0 \\
0 & \frac{1}{\ii \, \omega_{m} + \delta_{n}}
\end{array}\right), 
\end{eqnarray*}
where $\delta_{m n}$ is a Kronecker-delta in the NW position, we finally arrive at the Green's function for a SPW coupled to an array of MZMs, dependent on two spatial coordinates and Matsubara frequency and satisfying the Dyson equation:
\begin{eqnarray}
\label{eq:SPWGreenDyson1}
\hat{G}(x, x'; \ii \, \omega_{m}) & = & \hat{G}_{0}(x - x'; \ii \, \omega_{m}) \nonumber \\
 & + & \sum_{n} \hat{G}_{0}(x - n \, b; \ii \, \omega_{m}) \times \nonumber \\
& \times & \hat{V}_{n}(\ii \, \omega_{m}) \cdot \hat{G}(n \, b, x'; \ii \, \omega_{m}).
\end{eqnarray}
Here all the effects of the $n$-th topological NW are contained within the vertex:
\begin{equation}
\label{eq:Vertex}
\hat{V}_{n}(\ii \, \omega_{m}) = -\frac{\ii \, \omega_{m}}{2 (\omega^{2}_{m} + \delta^{2}_{n})} \, \left(\begin{array}{cc}
\vert \Gamma_{n} \vert^{2} & \Gamma^{2}_{n} \\
(\Gamma^{\ast}_{n})^{2} & \vert \Gamma_{n} \vert^{2}
\end{array}\right)\,.
\end{equation}
In Eq.\,(\ref{eq:SPWGreenDyson1}) $\hat{G}_{0}$ is the propagator in the SPW  given by:
\begin{eqnarray}
\label{eq:Ge0}
& \hat{G}_{0}(x-x'; \ii \, \omega_{m}) = \nonumber \\
& = \left(\begin{array}{cc}
g_{e}(x - x'; \ii \, \omega_{m}) & 0 \\
0 & g_{h}(x - x'; \ii \, \omega_{m}
\end{array}\right),
\end{eqnarray}
where the first diagonal element correspond to the free electron propagator in the SPW given by  $g_{e}( x - x' ; \ii \, \omega_{m}) = -\ii \, {\rm sgn}(\omega_{m})(m/k_{0}){\rm e}^{\ii \, {\rm sgn}(\omega_{m}) \, k_{0} \, \vert x-x' \vert}$  with $k_{0} = \sqrt{2 \, m \, (\mu_{w} + \ii \, \omega_{m})}$ and the second term to the  hole propagator in the SPW $g_{h}(x - x'; \ii \, \omega_m)=- g_{e}(x' - x;-\ii \, \omega_m)$.

An equation for $\hat{V}_{n} \cdot \hat{G}(n \, b, x')$ is obtained by multiplying \req{eq:SPWGreenDyson1} by $\hat{V}_{n}$ from the left, and setting $x = n \, b$. The solution of this equation is then expressed as:
\begin{eqnarray}
& \hat{V}_{n}(\ii \, \omega_{m}) \cdot \hat{G}(n \, b, x'; \ii \, \omega_{m}) = \nonumber \\
& = \sum_{l} \hat{T}_{n l}(\ii \, \omega_{m}) \cdot \hat{G}_{0}(l \, b - x'; \ii \, \omega_{m}), \label{eq:SPWDyson2}
\end{eqnarray}
where the matrix $\hat{T}_{n l}(\ii \, \omega_{m})$ plays the role of a kernel of the equation:
\begin{eqnarray}
\sum_{l} & \left[\delta_{n l} \hat{1} - \hat{V}_{n}(\ii \, \omega_{m}) \cdot (\hat{g}_{0})_{n l}(\ii \, \omega_{m}) \right] \times \nonumber \\
 \times & \hat{T}_{l m}(\ii \, \omega_{m}) = \hat{V}_{n}(\ii \, \omega_{m}) \, \delta_{n m}.\label{eq:TmatrixDef}
\end{eqnarray}
Here we introduce the shorthand notation $(\hat{g}_{0})_{n l}(\ii \, \omega_{m})$, which is the Green's function for a bare SPW evaluated at the discrete subset of positions of the MZMs, i.e.:
\begin{eqnarray}
& \left(\hat{g}_{0}\right)_{m n}(\ii \, \omega_{m}) = \left(\hat{g}_{0}\right)_{m - n}(\ii \, \omega_{m}), \nonumber \\
& \left(\hat{g}_{0}\right)_{n}(\ii \, \omega_{m}) = \hat{G}_{0}(n \, b; \ii \, \omega_{m}). \label{eq:DiscreteG0}
\end{eqnarray}

Finally, the solution of \req{eq:SPWGreenDyson1}, given \reqs{eq:SPWDyson2} and \rref{eq:TmatrixDef}, may be written as:
\begin{eqnarray}
\label{eq:SPWGreenDyson3}
\hat{G}(x, x'; \ii \, \omega_{m}) & = & \hat{G}_{0}(x - x'; \ii \, \omega_{m}) \nonumber \\
 & + & \sum_{n, m} \hat{G}_{0}(x - n \, b; \ii \, \omega_{m}) \times \nonumber \\
& \times & \hat{T}_{n m}(\ii \, \omega_{m}) \cdot \hat{G}_{0}(m \, b, x'; \ii \, \omega_{m}),
\end{eqnarray}
making it clear that $\hat{T}_{n m}(\ii \, \omega_{m})$ plays the role of a $T$-matrix for the SPW  with respect to the effect of the MZM array. The Green's function given by Eq.\,(\ref{eq:SPWGreenDyson3}) allows the calculation of the induced pair correlations in the SPW and is used for the analytical calculations in Sec.~\ref{sect3}. 

\subsection{\label{sec:ModelTB}Numerical tight-binding setup}
\begin{figure}[!t]
\includegraphics[width=0.6\linewidth]{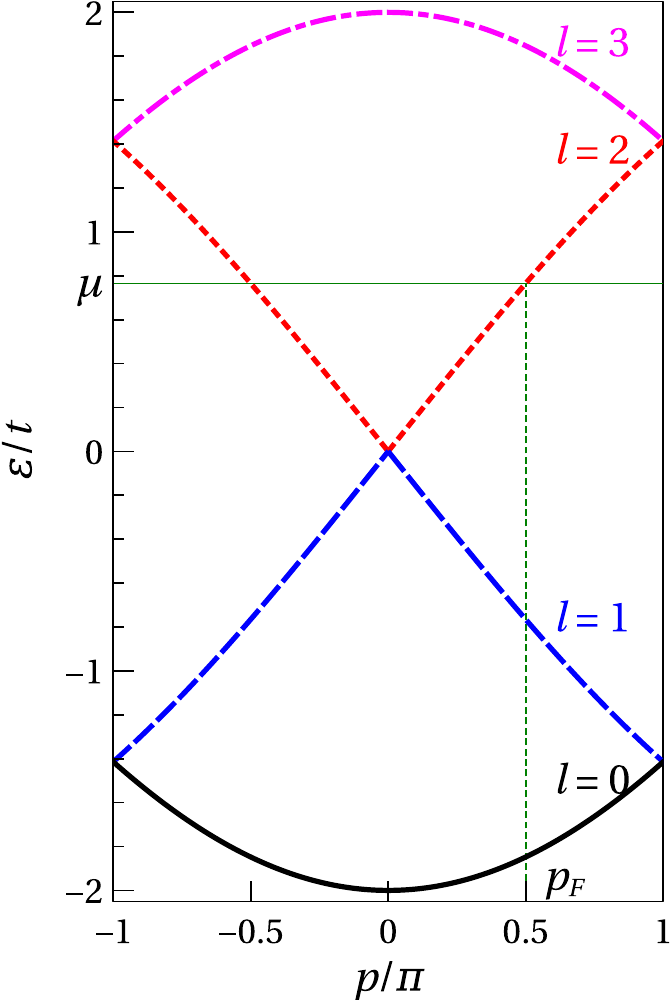}
\caption{\label{fig:TBFold}A sketch of a folded band dispersion with $z = 4$ bands. The chemical potential $\mu$  is chosen to cut the $l = 2$ conduction band (green line), with the (dimensionless) Fermi wavevector equal to $p_{\rm F} = 0.5 \, \pi$.}
\end{figure}

Beyond studying the MZM array using Green's functions we also adopt a tight-binding version of the model described by \reqs{eq:WireHamil}-\rref{eq:HybridizationHam} to facilitate numerical results going beyond the perturbative analytical treatment. Here we place $M$ MZMs on a ring with $z-1$ lattice sites of the SPW inbetween two MZMs and one SPW site at each MZM site, with the periodic boundary conditions $c_{n + z \, M} = c_{n}$ for the SPW modes and $f_{n + M} = f_{n}$ for the Bogoliubov modes in the $n$-th NW through which the MZM is expressed. The tight-binding version of \req{eq:WireHamil} is then given by:
\begin{equation}
 \mathcal{H}^{\rm TB}_{w} = -t \, \sum_{n = 0}^{z \, M - 1} \left\lbrace c^{\dagger}_{n} \, c_{n + 1} + \mathrm{H.c.} \right\rbrace  
- \mu \, \sum_{n = 0}^{z \, M - 1} c^{\dagger}_{n} \, c_{n}\,.
\end{equation}
For the clean, non-disordered system we can further impose periodic boundary conditions with period $z$, corresponding to a lattice constant equal to the MZM distance $b$. This leads to folding of the original SPW dispersion relation $\xi(p) = -2 t \, \cos(p/z) - \mu$, $z$ times as $p$ goes over from $-z \, \pi$ to $z \, \pi$ onto the new unit cell with Brillouin zone from $-\pi$ to $\pi$. Thus, specifying the band index $l$ ($0 \le l \le z - 1$) of the conduction band cut by the chemical potential, as well as the (dimensionless) Fermi wavevector $-\pi < p_{\rm F} \le \pi$, fixes the chemical potential $\mu$ to:
\begin{subequations}
\label{eq:Bands}
\begin{eqnarray}
& \mu = \mu_{l}(p_{\rm F}) = -2 t \, \cos\left( \frac{p_{\rm F} + 2 \pi \, m_{l}}{z}\right), \label{eq:ChemPotTB} \\
& m_{l} = \left\lbrace \begin{array}{lcl}
r &,& l = 2 \, r, \\
-r &,& l = 2 \, r - 1
\end{array}\right.. \label{eq:BandFoldTB}
\end{eqnarray}
\end{subequations}
The meaning of \reqs{eq:Bands} is sketched in Fig.~\ref{fig:TBFold}.

The tight-binding version of the tunnel Hamiltonian \req{TunnelHamil} becomes:
\begin{equation}
\label{eq:TunnelHamilTB}
\mathcal{H}^{\rm TB}_{\rm \Gamma}=\frac{\ii}{2 \sqrt{z}} \, \sum_{n = 0}^{M - 1}\left\lbrace \Gamma_{n}^{*} \gamma_{n} \, c_{z \, n}+ \Gamma_{n} \, \gamma_{n}c_{z \, n} \right\rbrace\,,
\end{equation}
where the factor $1/\sqrt{z}$ is necessary if we require $\vert \Gamma \vert$ to be the energy gap in the single-particle excitation spectrum in case of a periodic non-disordered MZM array $\Gamma_{n} = \Gamma$, just as in the continuum case.

We can now define a large $2 \, M \, (z + 1)$-dimensional Nambu spinor $\Psi = \left( c_{n}, c^{\dagger}_{n} \vert f_{m}, f^{\dagger}_{m} \right)^{\top}$, such that the tight-binding Hamiltonian may be written as a bilinear form $\mathcal{H}_{w} + \mathcal{H}_{\rm \Gamma} + \mathcal{H}_{\delta} = \frac{1}{2} \, \Psi^{\dagger} \cdot \check{H}_{\rm BdG} \cdot \Psi$, where the matrix $\check{H}_{\rm BdG}$ is the Bogoliubov-de Gennes (BdG) Hamiltonian. Diagonalizing this matrix generates the eigenvalues $E_{N}$ and eigenvectors $u^{(N)}_{a}$, and all single-particle correlators are then in principle calculable. In particular, the tight-binding version of Eq.\,(\ref{eq:GreenFuncMat}) is the upper left $z \, M \times z \, M$ block of the inverse matrix $\check{G}_{\rm BdG}(\ii \, \omega_{m}) = \left(\ii \, \omega_{m} \, \check{1} - \check{H}_{\rm BdG}\right)^{-1}$, which using the eigenvalues and eigenvectors can be calculated as:
\begin{equation}
\check{G}_{\rm BdG}(\ii \, \omega_{m}) = \sum_{N} u^{(N)}_{a} \, \left( u^{(N)}_{b} \right)^{\ast} \frac{1}{\ii \, \omega_{m} - E_{N}}. \label{eq:BdGGreen}
\end{equation}
From this we can directly extract the even- and odd-$\omega$ anomalous components $G^{\rm E, O}_{e h}$ as well as the LDOS $N_{i}(\varepsilon)$ using:
\begin{subequations}
\label{eq:BdGExpressions}
\begin{eqnarray}
& \left[G^{\rm E}_{e h}\right]_{i j}(\ii \, \omega_{m}) = -\sum_{N} u^{(N)}_{i} \Big[u^{(N)}_{j + z \, M} \Big]^{\ast}  \frac{E_{N}}{\omega^{2}_{m} + E^{2}_{N}}, \label{eq:GehEvenTB} 
\\
& \left[G^{\rm O}_{e h}\right]_{i j}(\ii \, \omega_{m}) = -\sum_{N} u^{(N)}_{i} \Big[u^{(N)}_{j + z \, M} \Big]^{\ast}  \frac{\ii \, \omega_{m}}{\omega^{2}_{m} + E^{2}_{N}}, \label{eq:GehOddTB} 
 \\
& N_{i}(\varepsilon) = \sum_{N} u^{(N)}_{i} \, \Big[u^{(N)}_{j} \Big]^{\ast} \frac{\eta}{\pi \, \big[ (\varepsilon - E_{N} )^{2} + \eta^{2}\big]}, \label{eq:LDOSTB}
\end{eqnarray}
\end{subequations}
where $\eta$ is a small positive smearing factor of a Dirac delta function.


\section{\label{sect3}Analytical Results}
In this part we demonstrate that disorder in the phases of the complex couplings are detrimental for the induced pair amplitudes. Moreover, we show that when there is a finite relative phase  between the couplings of at least two MZMs, bound states are induced in the SPW.
In order to show these arguments, we consider three simple, yet useful, cases that provide a clear visualization of the role of complex couplings on the pair amplitudes.
 
\subsection{\label{sec:ResultsSingleMZM}SPW coupled to a single MZM}
We start by considering the simplest case, a single MZM coupled to the SPW. In this case the system of equations \req{eq:TmatrixDef} is actually a single equation involving $2\times2$ matrices in Nambu space, which is readily solved:
\begin{equation}
\label{eq:1MZM}
 \hat{T}(\ii \, \omega_{m})  = -\frac{\ii \, \omega_{m}}{2 \, D} \, \begin{pmatrix}
 \vert \Gamma \vert{2} & \Gamma^{2} \\
\left(\Gamma^{\ast}\right)^{2} & \vert \Gamma \vert^{2}
\end{pmatrix}, 
\end{equation}
where $D = \omega^{2}_{m} + \delta^{2}   
+ \frac{\ii \, \omega_{m} \, \left[g_{e}(0; \ii \, \omega_{m}) + g_{h}(0; \ii \, \omega_{m}) \right]}{2} \, \, \vert \Gamma \vert^{2}$ is an even function of frequency $\omega_{m}$. 
The only difference between Eq.\,(\ref{eq:1MZM}) and \req{eq:Vertex} is thus the substitution of the denominator $D$.
This clarifies the meaning of the $T$-matrix as involving a series of coherent events between the MZM and the SPW of scattering as an electron and back-scattering as a hole. Then, by using \req{eq:SPWGreenDyson3}, the Green's function for the SPW is easily obtained, with the entries specified in Eq.~\req{eq:GreenFuncMat} given by:
\begin{equation}
\label{G1}
\begin{split}
& G_{ee}(x,x',\ii \, \omega_{m}) = g_{e}(x-x', \ii \, \omega_{m})\\
& + g_{e}(x,0,\ii \, \omega_{m}) \, T_{ee} \, g_{e}(0,x',\ii \, \omega_{m}), \\
& G_{eh}(x,x',\ii \, \omega_{m}) = g_{e}(x,0,\ii \, \omega_{m}) \, T_{eh} \, g_{h}(0,x',\ii \, \omega_{m}), \\
& G_{he}(x,x',\ii \, \omega_{m}) = g_{h}(x,0,\ii \, \omega_{m}) \, T_{he} \, g_{e}(0,x',\ii \, \omega_{m}), \\
& G_{hh}(x,x',\ii \, \omega_{m}) = g_{h}(x-x',\ii \, \omega_{m}) \\
&+ g_{h}(x,0,\ii \, \omega_{m}) \, T_{hh} \, g_{h}(0,x',\ii \omega_{m})\,,
\end{split}
\end{equation}
where  $T_{ij}$ are the elements of the $\hat{T}$ matrix given by Eq.\,(\ref{eq:1MZM}). 

The diagonal (normal) elements of the Green's function allow the calculation of the LDOS, while from the off-diagonal (anomalous) terms we extract information about the pair correlations induced in the SPW. Since we are interested in the induced pair correlations, we now discuss $G_{eh}$ which written out is:
\begin{equation}
\label{GehSMZM}
G_{eh}(x,x',\ii \, \omega_{m})= -\frac{m^{2} \, T_{e h}}{ \vert k_{0} \vert^{2}}\,{\rm e}^{\ii\,{\rm sgn}(\omega_{m}) \, \left[k_{0} \, \vert x \vert -k_{0}^{\ast} \,  \vert x' \vert\right]},
\end{equation}
where we have used the expressions for the free particle propagators in the SPW given in Eq.\,(\ref{eq:Ge0}), $T_{eh}=-(\ii \, \omega_{m}/D)\Gamma^{2}$, and $k_{0}$ is the electron wave vector defined after Eq.\,\ref{eq:Ge0}. The pair amplitude $G_{eh}$ thus exhibits translational invariance breaking through the exponent $k_{0} \, \vert x \vert -k_{0}^{\ast} \, \vert x' \vert$, which mixes electron ($k_{0}$) and hole ($k_{0}^{\ast}$)  wavevectors with different spatial 
coordinates. This is similar to the breaking of spatial parity in other superconducting junctions.\cite{Nagaosa12,PhysRevB.96.155426,PhysRevB.98.075425} In order to visualize this point it is useful to write the wave vector $k_{0}$, defined after Eq.\,(\ref{eq:Ge0}), in the limit $\vert \omega_{m} \vert \ll \mu$, where $k_{0} \approx k_{\rm F}+\ii \, \kappa$. Rearranging the exponent of \req{GehSMZM}, results in $k_{\rm F} \, (\vert x \vert - \vert x'\vert)+ \ii\,\kappa \, ( \vert x \vert + \vert x' \vert)$ with $\kappa=(\omega_{m} \, k_{\rm F})/(2\mu)$ and $k_{\rm F}=\sqrt{2m \, \mu}$ (we choose a system of units in which $\hbar = 1$). Plugging this expression into \req{GehSMZM}, leads to an exponential decay term with decay length given by the inverse of $\kappa$ and an oscillatory term governed by the Fermi wave-vector. This oscillatory term can further be written in terms of cosines and sines, which are even and odd-functions in space, respectively, and thus leads to the coexistence of even- and odd-$\omega$ components of the pair amplitude.\cite{PhysRevB.96.155426,PhysRevB.98.075425} Interestingly, locally in space, i.e.~$x=x'$, $G_{eh}$ is purely odd in frequency, an induced effect that comes directly from the MZM. The odd-$\omega$ dependence is here captured in the element $T_{eh}$, since its denominator $D$ is an even function of $\omega_{m}$.
We  therefore conclude that the induced local pairing due to a single MZM is odd in frequency, in agreement with previously reported the odd-$\omega$ nature of Majorana pair amplitudes.\cite{PhysRevB.87.104513,PhysRevB.92.121404,Lee16,PhysRevB.96.155426,PhysRevB.97.134523,Tanaka19,PhysRevB.99.184512,thanos2019} 

Furthermore, from Eqs.\,(\ref{G1}) we observe that the anomalous elements are proportional to $\Gamma^{2}$. This stems from Eq.\,(\ref{eq:1MZM}), where the diagonal  elements $T_{ee,hh}$ are proportional to $ \vert \Gamma \vert^{2}$, while the off-diagonal  $T_{eh}$ is proportional to $\Gamma^{2}$. This is in notable contrast to the findings of Ref.\,\onlinecite{Lee16}, where the pair amplitude was assumed to always be proportional to real couplings.
If $\Gamma$ is real, it does not present any issues for the pair amplitudes, however, it does when $\Gamma$ is complex. In fact, by performing a disorder average over the phase of $\Gamma$, the expression for $G_{eh}$ might give zero or be exponentially suppressed, depending on the disorder distribution. Of course, there is no physical meaning  of performing disorder average over the phase of $\Gamma$ when only a single MZM is coupled to the SPW. However, this simple example still clearly shows the potentially detrimental effect of disorder averaging over the phases of $\Gamma$ on the pairing amplitudes. As we will see in the following subsections, this seemingly innocent effect at this level has profound consequences when considering more MZMs coupled to the SPW.

\subsection{\label{sec:Results2MZM}SPW coupled to two MZMs}
The simplest system where the relative phase of the coupling strengths $\Gamma$ has a physical manifestation is a SPW coupled to $2$ MZMs with distance $b$ and complex couplings $\Gamma_{1,2}$. Then, \req{eq:TmatrixDef} is a $4\times 4$ matrix in site and Nambu spaces, but still analytically solvable. Following the same method as for a single MZM we obtain the even- and odd-$\omega$ pair amplitudes
\begin{widetext}
\begin{subequations}
\label{EQ2MZMs}
\begin{eqnarray}
& G_{eh}^{\rm O}(x,x';\ii\, \omega_{m}) = -\frac{m^{2}}{\vert k_{0} \vert^{2}} \left\lbrace
2 \, T_{11} \, {\rm cos}[s \, k_{\rm F}(\vert x \vert - \vert x' \vert)] \, {\rm e}^{-s \, \kappa ( \vert x \vert + \vert x' \vert)} \right. \nonumber \\
& + 2 \, T_{22} \, {\rm cos}[s \, k_{\rm F}( \vert x - b \vert - \vert b -x' \vert)] \, {\rm e}^{-s \, \kappa(\vert x- b \vert + \vert b -x' \vert)} \nonumber \\
& +{\rm e}^{-s \, \kappa( \vert x - b \vert + \vert x' \vert )} \, \left[ T_{21}{\rm e}^{\ii \, s \, k_{\rm F}( \vert x - b \vert -\vert x' \vert)} + T_{12} \, {\rm e}^{-\ii \, s \, k_{\rm F}(\vert x- b \vert - \vert x' \vert)}\right] \nonumber \\
& \left. + {\rm e}^{-s\kappa( \vert x \vert + \vert x'- b \vert )} \, \left[ T_{12}{\rm e}^{\ii \, s \, k_{\rm F}( \vert x \vert - \vert x'- b \vert )} + T_{21} \, {\rm e}^{-\ii \, s \, k_{\rm F} \, (\vert x \vert - \vert x' - b \vert )}\right]
\right\rbrace, \label{EQ2MZM2Odd} \\
& G_{eh}^{\rm E}(x,x';\ii\, \omega_{m}) = -\frac{m^{2}}{\vert k_{0} \vert^{2}} \left\lbrace
2 \ii \, T_{11} \, {\rm sin}[s \, k_{\rm F}(\vert x \vert - \vert x' \vert)] \, {\rm e}^{-s \, \kappa ( \vert x \vert + \vert x' \vert)} \right. \nonumber \\
& + 2 \ii \, T_{22} \, {\rm sin}[s \, k_{\rm F}( \vert x - b \vert - \vert b -x' \vert)] \, {\rm e}^{-s \, \kappa(\vert x- b \vert + \vert b -x' \vert)} \nonumber \\
& +{\rm e}^{-s \, \kappa( \vert x - b \vert + \vert x' \vert )} \, \left[ T_{21}{\rm e}^{\ii \, s \, k_{\rm F}( \vert x - b \vert -\vert x' \vert)} - T_{12} \, {\rm e}^{-\ii \, s \, k_{\rm F}(\vert x- b \vert - \vert x' \vert)}\right] \nonumber \\
& \left. + {\rm e}^{-s\kappa( \vert x \vert + \vert x'- b \vert )} \, \left[ T_{12}{\rm e}^{\ii \, s \, k_{\rm F}( \vert x \vert - \vert x'- b \vert )} - T_{21} \, {\rm e}^{-\ii \, s \, k_{\rm F} \, (\vert x \vert - \vert x' - b \vert )}\right]
\right\rbrace, \label{EQ2MZM2Even}
\end{eqnarray}
\end{subequations}
where  $s={\rm sign}(\omega_{m})$, $\kappa=(\omega_{m}k_{\rm F})/(2\mu)$, $k_{\rm F}=\sqrt{2m\mu}$, and $T_{ij}=N_{ij}^{eh}/\bar{D}$ where the numerator are the $eh$ components
\begin{subequations}
\label{eq:EHNum2MZM}
\begin{eqnarray}
& N^{eh}_{11}(\ii \, \omega_{m}) = -\frac{\ii \, \omega_{m} \, \Gamma_{1}}{2} \, \left[ \Gamma_{1} \, g_{h}(0; \ii \, \omega_{m}) \, D_{2}(\ii \, \omega_{m}) + \Gamma_{2} \, g_{h}(b; \ii \, \omega_{m}) \, \tilde{K}(\ii \, \omega_{m}) \right], \label{eq:N11EH2MZM} \\
& N^{eh}_{12}(\ii \, \omega_{m}) = -\frac{\ii \, \omega_{m} \, \Gamma_{1}}{2} \, \left[ \Gamma_{1} \, g_{h}(-b; \ii \, \omega_{m}) \, D_{2}(\ii \, \omega_{m}) + \Gamma_{2} \, g_{h}(0; \ii \, \omega_{m}) \, \tilde{K}(\ii \, \omega_{m}) \right], \label{eq:N12EH2MZM} \\
& N^{eh}_{21}(\ii \, \omega_{m}) = -\frac{\ii \, \omega_{m} \, \Gamma_{2}}{2} \, \left[ \Gamma_{1} \, g_{h}(0; \ii \, \omega_{m}) \, K(\ii \, \omega_{m}) + \Gamma_{2} \, g_{h}(b; \ii \, \omega_{m}) \, D_{1}(\ii \, \omega_{m}) \right], \label{eq:N21EH2MZM} \\
& N^{eh}_{22}(\ii \, \omega_{m}) = -\frac{\ii \, \omega_{m} \, \Gamma_{2}}{2} \, \left[ \Gamma_{1} \, g_{h}(-b; \ii \, \omega_{m}) \, K(\ii \, \omega_{m}) + \Gamma_{2} \, g_{h}(0; \ii \, \omega_{m}) \, D_{1}(\ii \, \omega_{m}) \right]\,,\label{eq:N22EH2MZM}
\end{eqnarray}
\end{subequations}
\end{widetext}
where  $g_{e,h}$ correspond to the free electron and hole propagators in the SPW defined in Eq.\,(\ref{eq:Ge0}).
Moreover, the denominator $\bar{D}$ is given by
$\bar{D}(\ii \, \omega_{m}) = D_{1}(\ii \, \omega_{m}) \, D_{2}(\ii \, \omega_{m}) - K(\ii \, \omega_{m}) \, \tilde{K}(\ii \, \omega_{m}),$
where $D_{i}(\ii \, \omega_{m})$ correspond to the denominator of the expression in \req{eq:1MZM} but for each MZM separately ($i = 1, 2$), and
$
K(\ii \, \omega_{m}) = -\frac{\ii \, \omega_{m}}{2}\, \left[ \Gamma_{1} \, \Gamma^{\ast}_{2} \, g_{e}(b; \ii \, \omega_{m}) + \Gamma^{\ast}_{1} \, \Gamma_{2} \, g_{h}(b; \ii \, \omega_{m}) \right],\, 
\tilde{K}(\ii \, \omega_{m}) = -\frac{\ii \, \omega_{m}}{2} \, \left[ \Gamma_{1} \, \Gamma^{\ast}_{2} \, g_{h}(-b; \ii \, \omega_{m}) + \Gamma^{\ast}_{1} \, \Gamma_{2} \, g_{e}(-b; \ii \, \omega_{m}) \right].$ 

Based on these results we first conclude from Eqs.\,(\ref{EQ2MZMs}) that both even- and odd-$\omega$ pair correlations are induced in the SPW due to its coupling to MZMs. Both pair amplitudes also acquire a dependence on the phases of the complex couplings $\Gamma_{1,2}$, as evident in Eqs.\,(\ref{eq:EHNum2MZM}).
Moreover, since these amplitudes correspond to equal-spin spin-triplet pairing, the even-frequency component is finite non-locally but vanishes locally, which is similar to the even-$\omega$ pairing discussed for the single MZM in the previous subsection. 
On the other hand, we find that odd-$\omega$ pairing exists both locally and non-locally. Both even-$\omega$ and odd-$\omega$ pair amplitudes exponentially decay from the position of the MZMs with a decay length determined by the inverse of $\kappa$, specified in the previous subsection.
Second, we conclude that the pair amplitudes given by Eqs.\,(\ref{EQ2MZMs}) emerge proportional to $\Gamma_{1(2)}^{2}$ or $\Gamma_{1}\Gamma_{2}$ through the numerator of the $T_{ij}$, $N_{ij}^{eh}$, given by Eqs.\,(\ref{eq:EHNum2MZM}). This, therefore, implies that any average over the complex phases of $\Gamma_{1,2}$  leads to a suppression of the induced pair correlations in the SPW, demonstrating a detrimental role of complex couplings. 

\begin{figure}[!b]
\includegraphics[width=0.95\linewidth]{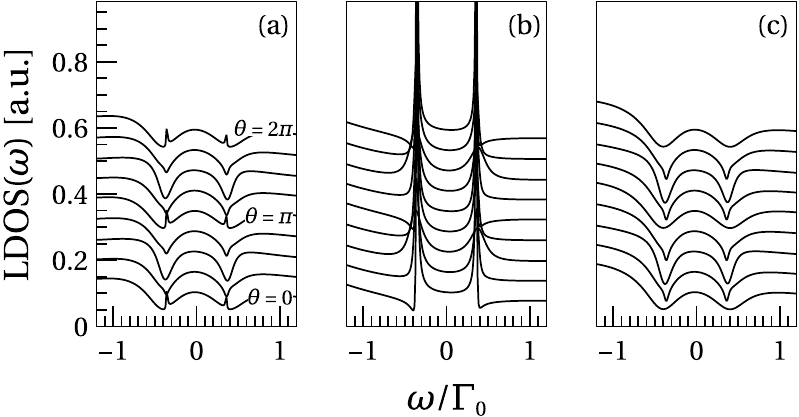}
\caption{\label{fig:2MZMLDOSb=1}LDOS (in arbitraty units) in the SPW for two MZMs as a function of energy. Each curve is shifted vertically for illustrative purposes and corresponds to an equidistant value of the relative phase $\theta$ (denoted only for $\theta = 0$, $\pi$ and $2\pi$ in panel (a) for orientation) of the coupling to two MZMs between $0$ and $2\pi$ and obtained under the first MZM (a), at $1/4$ (b), and $1/2$ (c) of the MZM distance $b$ from the first MZM. Here, coupling strength for both MZMs is $\Gamma_{0} = 0.4 \, \varepsilon_{\rm F}$, $b = \lambda_{\rm F}$, and the hybridization $\delta = 0.4 \, \Gamma_{0}$ for both MZMs. The smearing factor is set to $\eta = 0.5 \times 10^{-4} \, \Gamma_{0}$.}
\end{figure}

Before concluding this part, we explicitly demonstrate the effect of a finite relative phase $\theta$ between the couplings of the two MZMs $\Gamma_{1,2}$. This is illustrated in Fig.~\ref{fig:2MZMLDOSb=1}, where we plot the LDOS in the SPW as a function of frequency for nine values of the phase $\theta$ uniformly distributed on the segment $[0, 2\pi]$. We calculate the LDOS  from the standard expression ${\rm LDOS}(x,\omega)=(-1/\pi){\rm Im} G_{ee}(x,x;\omega+i\eta)$, with $\omega$  real frequency, $\eta$ an infinitesimal positive number, and $G_{ee}$ the electron-electron Green's function obtained after solving \req{eq:SPWGreenDyson3}.
The system is symmetric with respect to the midpoint $x = b/2$ between the two MZMs and we therefore plot the LDOS at three representative points on the side closer to the first MZM, where we measure the MZM distance $b$ in units of the Fermi wavelength $\lambda_{\rm F} = 2\pi/k_{\rm F}$ in the SPW.

The most prominent feature in Fig.~\ref{fig:2MZMLDOSb=1} is the presence of resonance-like peaks at $\pm \delta$, corresponding to the leaking of the MZM states into the SPW. The weight of these peaks is highly dependent on position, having the highest weight (anti node) at $b/4$, while the lowest weight (node) is directly under the MZM and also in the middle between the two MZMs at $b/2$. These correspond to delocalized states in the SPW with a half-wavelength (node-to-node distance) $b/2$, i.e.~a wavelength equal to $b$. Since in this plot we have chosen $b = \lambda_{\rm F}$, it is not clear from these results which of these length scales determines the wavelength of the bound state like feature. We therefore investigate other MZM separations in App.~\ref{sec:AppMZMBound} and find that the wavelength is determined by $\lambda_{\rm F}$. In addition, it is clear that the spectrum is symmetric in $\theta$ relative to $\theta = \pi$. The peaks are sharpest for $\theta = 0$, $\pi$, and $2\pi$, and have the largest smearing for $\theta = \pi/2, 3 \pi/2$. The smearing, associated with a finite lifetime of these resonance states, is highly phase-dependent, which is reminiscent of the conditions for constructive and destructive interference. To summarize, these results suggest that delocalized bound states appear naturally systems with multiple MZMs due to the complex couplings, but that their behavior is also naturally expected to be less controllable as the number of Majorana NWs increases.


\subsection{\label{sec:ResultsArraySelfEn}SPW coupled to a MZM array}
Having studied one and two MZM coupled to the SPW, we finally turn to the case of an infinite array of MZMs and for this system calculate the induced pair amplitudes  in the SPW using disorder-averaging to capture the intrinsic variability in the complex phases of the MZM couplings.

For an infinite array, it is convenient to express \req{eq:TmatrixDef} in terms of the Fourier series coefficients:
\begin{subequations}
\label{eq:FourierDefs}
\begin{eqnarray}
& \hat{\tilde{T}}(p, p'; \ii \, \omega_{m}) = \nonumber \\
& = \sum_{n, m = -\infty}^{\infty} \hat{T}_{n m}(\ii \, \omega_{m}) \, \ee^{-\ii \, (p \, n - p' \, m)}, \label{eq:TmatrixFourier} \\
& \hat{\tilde{V}}(p; \ii \, \omega_{m}) = \sum_{n = -\infty}^{\infty} \hat{V}_{n}(\ii \, \omega_{m}) \, \ee^{-\ii p \, n}, \label{eq:VmatrixFourier} \\
& \hat{\tilde{g}}_{0}(p; \ii \, \omega_{m}) = \sum_{n = -\infty}^{\infty} (\hat{g}_{0})_{n}(\ii \, \omega_{m}) \, \ee^{-\ii \, p \, n}, \label{eq:g0Fourier}
\end{eqnarray}
\end{subequations}
where $p, p'$ are continuous variables on the interval $(-\pi, \pi]$. The functions defined in \reqs{eq:FourierDefs} are all periodic in each $p$ with a period $2\pi$. With this \req{eq:TmatrixDef} acquires the form:
\begin{eqnarray}
& \hat{\tilde{T}}(p, p'; \ii \, \omega_{m}) = \hat{\tilde{V}}(p - p'; \ii \, \omega_{m}) \nonumber \\
& + \int_{-\pi}^{\pi} \frac{d p''}{2\pi} \, \hat{\tilde{V}}(p - p''; \ii \, \omega_{m}) \times \nonumber \\
& \times \hat{\tilde{g}}_{0}(p''; \ii \, \omega_{m}) \cdot \hat{\tilde{T}}(p'', p'; \ii \omega_{m}). \label{eq:TmatFourierEq} 
\end{eqnarray}
This is an integral equation for $\hat{\tilde{T}}(p, p';\ii \, \omega_{m})$ as a function of continuous variables, which reflects the now infinite dimension of the matrix equation \req{eq:TmatrixDef}. Nevertheless, it is a convenient formulation for three reasons.

First, the periodicity with a lattice constant $b$ imposed on the continuum wire by the MZM array is reflected in the folded $\hat{\tilde{g}}_{0}(p; \ii \, \omega_{m})$. Expressing it in terms of the continuum Green's function $\hat{\tilde{G}}_{0}(k, \ii \, \omega_{m})$, the following relation holds:
\begin{equation}
\label{eq:DiscreteG0Fourier}
\hat{\tilde{g}}_{0}(p; \ii \, \omega_{m}) = \frac{1}{b} \, \sum_{m = -\infty}^{\infty} \hat{\tilde{G}}_{0}\left(\frac{p + 2 \pi \, m}{b}; \ii \, \omega_{m} \right).
\end{equation}
In fact, due to the simple parabolic dispersion relation, the sum over $m$ in \req{eq:DiscreteG0Fourier} may be performed analytically with the result:
\begin{subequations}
\label{eq:FourierBareSPWAnalytic}
\begin{eqnarray}
& \tilde{g}_{e}(p; \ii \, \omega_{m}) = -\frac{m}{2 k_{0}}
\left\lbrace {\rm cot} \left(\frac{p-k_{0} \, b}{2}\right)-{\rm cot}\left(\frac{p + k_{0} \, b}{2}\right) \right. \nonumber \\
& \left. + 2 \ii \, {\rm Floor} \left[\frac{{\rm arg}(p + k_{0} \, b)}{2\pi}\right]\right\rbrace, \label{eq:ElectronBareSPWAnalytic} \\
& \tilde{g}_{h}(p; \ii \, \omega_{m}) = \frac{m}{2 k^{\ast}_{0}}
\left\lbrace {\rm cot}\left(\frac{p - k_{0}^{\ast} \, b}{2}\right) - {\rm cot}\left(\frac{p + k^{\ast}_{0} \, b}{2}\right)\right. \nonumber \\
& \left. + 2 \ii \, {\rm Floor}\left[\frac{{\rm arg}(p + k_{0}^{\ast} \, b)}{2\pi}\right]
\right\rbrace, \label{eq:HoleBareSPWAnalytic}
\end{eqnarray}
\end{subequations}
where $k_{0}$ is the electron wavevector defined after Eq.\,(\ref{eq:Ge0}).

Second, the Fourier-transformed vertex coefficients $\hat{\tilde{V}}(p - p';\ii \, \omega_{m})$ become approximately normally distributed when disorder is taken into account, regardless of the exact distribution of the real-space $\hat{V}_{n}(\ii \, \omega_{m})$ according to the Central Limit Theorem. This allows us to apply Wick's theorem for an average of a product of several such terms in an expansion.

Third, after disorder averaging, the $T$-matrix recovers invariance under translations by a lattice constant $b$, i.e. $\avg{\hat{T}}_{n m}(\ii \, \omega_{m}) = \avg{\hat{T}}_{n + k, m + k}(\ii \, \omega_{m})$. This implies that the Fourier transform $\avg{\hat{\tilde{T}}}(p, p'; \ii \, \omega_{m})$ defined by \req{eq:FourierDefs} becomes diagonal in $p$, $p'$:
\begin{equation}
\label{eq:TmatrixDisAvg}
\avg{\hat{\tilde{T}}}(p, p'; \ii \, \omega_{m}) = 2 \pi \, \delta(p - p') \, \avg{\hat{\tilde{T}}}(p; \ii \, \omega_{m}).
\end{equation}

Using the Fourier series formulation we can now, from \reqs{eq:SPWGreenDyson3}, \rref{eq:TmatrixFourier}, \rref{eq:g0Fourier}, and \rref{eq:TmatrixDisAvg}, express the disorder-averaged Green's function for the SPW as:
\begin{eqnarray}
\label{eq:SPWGreenDisorderAvg}
& \avg{\hat{G}}(x, x'; \ii \, \omega_{m}) = \int_{-\infty}^{\infty} \frac{d k}{2 \pi} \hat{\tilde{G}}_{0}(k; \ii \, \omega_{m}) \, \ee^{\ii \, k \, (x - x')} \nonumber \\
& + \int_{-\pi}^{\pi}\frac{d p}{2 \pi} \, \ee^{\ii \, p \, \frac{x - x'}{b}} \nonumber \\
& \times \, \hat{\tilde{g}}_{0}(p; \ii \, \omega_{m}) \cdot \avg{\hat{\tilde{T}}}(p; \ii \, \omega_{m}) \cdot \hat{\tilde{g}}_{0}(p; \ii \, \omega_{m}).
\end{eqnarray}
Here we note that the second term incorporates the effects of BZ folding through the matrices $\hat{\tilde{g}}_{0}(p; \ii \, \omega_{m})$, with integration over $p$ only in the interval from $-\pi$ to $\pi$. As a direct consequence, the effects of disorder averaging are now completely encompassed within the disorder-averaged $\avg{\hat{\tilde{T}}}$-matrix. Finding a general expression for the disorder-averaged $\avg{\hat{\tilde{T}}}$-matrix is straightforward but lengthy and we therefore provide the details as Appendix \ref{sec:AppDisAvg}.

We are here primarily interested in the induced anomalous correlations in the SPW, characterized by the electron-hole component of the Green's function. To obtain them, according to Eq.\,(\ref{eq:SPWGreenDisorderAvg}) and the fact that both $\hat{\tilde{G}}_{0}(k; \ii \, \omega_{m})$ and $\hat{\tilde{g}}_{0}(p; \ii \, \omega_{m})$ are diagonal in Nambu space, only the electron-hole component of the disorder-averaged $\avg{\tilde{T}^{eh}}$ is needed. The detailed derivation of expressions for  $\avg{\hat{\tilde{T}}}$  within the  second Born approximation (i.e.~keeping only the $n = 1$ term in \req{eq:1PIselfenergy}, which is of second order in $\delta \hat{\tilde{V}}$) are provided in Appendix~\ref{sec:AppSelfEn}, and here we produce schematic, but, nevertheless sufficiently detailed expressions to draw the most important conclusions. In particular, for the electron-hole component we obtain
\begin{widetext}
\begin{equation}
\label{arrayanalyticEq}
\begin{split}
\avg{\tilde{T}^{eh}}(p; \ii \, \omega_{m}) &= \frac{1}{D_{T}(p; \ii \, \omega_{m})} \, \left\lbrace \phi_{1}(\ii \, \omega_{m}) \, B_{1} + \phi_{2}(\ii \, \omega_{m}) \left[A_{2} \, g^{eh}_{1}(\ii \, \omega_{m}) + C_{2} \, g^{he}_{1} (\ii \, \omega_{m}) \right.\right.  \\
& \left. \left. + B_{2} \, (g^{ee}_{1}(\ii \, \omega_m) + g^{hh}_{1}(\ii \, \omega_{m})) \right] - \mathcal{F}\right\}\\
 \mathcal{F}&=\frac{B_{1} \, \phi^{2}_{1}(\ii \, \omega_m) \, \left[ A_{1} \, (\tilde{g}_{e}(p; \ii \, \omega_{m}) + \tilde{g}_{h}(p; \ii \, \omega_{m})) - \phi_{1}(\ii \, \omega_{m}) \, (A^{2}_{1} - \vert B_{1} \vert^{2}) \, \tilde{g}_{e}(p; \ii \, \omega_{m}) \, \tilde{g}_{h}(p; \ii \, \omega_{m}) \right]}{1 - \phi_{1}(\ii \, \omega_{m}) \, A_{1} \left( \tilde{g}_{e}(p; \ii \, \omega_{m}) + \tilde{g}_{h}(p; \ii \, \omega_{m}) \right) + \phi^{2}_{1}(\ii \, \omega_{m}) \, \left( A^{2}_{1} - \vert B_{1} \vert^{2} \right) \, \tilde{g}_{e}(p; \ii \, \omega_{m}) \, \tilde{g}_{h}(p; \ii \, \omega_{m})} , 
\end{split}
\end{equation}
\end{widetext}
where $\tilde{g}_{e,h}$ are given by Eqs.\,(\ref{eq:FourierBareSPWAnalytic}).
The denominator $D_{T}$ is here an even function of $\omega_{m}$, but its explicit form is not necessary for our discussion. Similarly, the exact form of $\phi_{1}(\ii \, \omega_{m})$, given by \req{eq:DeltaAv1}, and $\hat{g}_{1}(\ii \, \omega_{m})$, given by \reqs{eq:g1def}, \rref{eq:g1matrix}, is not important for our discussion.

Although the expression for the disorder-averaged pair amplitude $\avg{G^{eh}}$ that follows from combining \reqs{eq:SPWGreenDisorderAvg} and \rref{arrayanalyticEq} is not simple, some conclusions can be drawn already from analyzing the components of $\avg{\tilde{T}^{eh}}$. For example, we obtain that  $A_{2}= \avg{|\Gamma|^{4}}$, $B_{2}= \avg{|\Gamma|^{2}\Gamma^{2}}$, and $C_{2}= \avg{\Gamma^{4}}$, while $A_{1}= \avg{ \vert\Gamma \vert^{2}}$ and $B_{1}= \avg{\Gamma^{2}}$, see Appendix \ref{sec:AppSelfEn} for details on their evaluation. In the generic situation, we expect $D_{\phi}$ to be close to $1$. Phase disorder-averaging thus forces $C_{2} \ll B_{2} \ll A_{2}$. This directly results in the  $A_{2}$-term being the dominant term in the square brackets expression in \req{arrayanalyticEq}. 
However, this term also contains a factor of $\avg{g^{eh}_{1}}$ that, according to \reqs{eq:g1def} and \rref{eq:FirstMoment} is also proportional to $B_{1}$. Thus, even this dominating term is in fact highly suppressed after phase-disorder averaging, because $B_{1} \ll A_{1}$. As a consequence, the disorder-averaged pair amplitude in Eq.\,(\ref{arrayanalyticEq}) is small, and we can conclude that both the even- and odd-$\omega$ induced pair correlations is highly suppressed in an MZM array.

Additionally, we have verified that the formation of bound states due to the coupling between the SPW and the array of MZMs follows the case discussed in previous section for two MZMs. In fact, we find bound states emerging at energies around the Majorana splittings $\delta_{n}$. However, unlike the case of two MZMs, where the two isolated modes are not able to open a gap in the continuum of the SPW spectrum, in the case of an array such a gap is opened, and the in-gap states also do not acquire smearing, but, are shifted in energy instead. We briefly discuss  this feature in the following section.

We conclude this analytical part of our work by summarizing our findings: A MZM coupled to a SPW induces even- and odd-$\omega$ correlations in the SPW, whose amplitude is proportional to $\Gamma^{2}$. Locally in space only odd-$\omega$ pairing exists. When two MZMs are coupled to the SPW with different complex couplings, delocalized bound states emerge in the SPW between the two MZMs, and both pair amplitudes become dependent on the phase difference and proportional to $\Gamma_{1(2)}^{2}$ or $\Gamma_{1}\Gamma_{2}$. This already on the level of two MZMs indicate a suppression of the induced pairing in the presence of phase disorder in couplings $\Gamma_i$. Finally, for an infinite array, we verify that the induced pair amplitudes in the SPW are necessarily considerably suppressed under disorder-averaging over the phases of the complex couplings.

\section{\label{sec:ResultsNumerics}Numerical Results}
In order to demonstrate the effects of phase-disorder, without relying on the approximations that were essential in obtaining closed-form expressions in the analytical approach in Sec.~\ref{sec:ResultsArraySelfEn}, such as the Born approximation for the infinite array, and also going beyond a qualitative interpretation of disorder effects, 
we next perform a numerical diagonalization of a tight-binding model described in Sec.~\ref{sec:ModelTB} and add disorder explicitly.

\begin{figure*}[!th]
\includegraphics[width=0.7\linewidth]{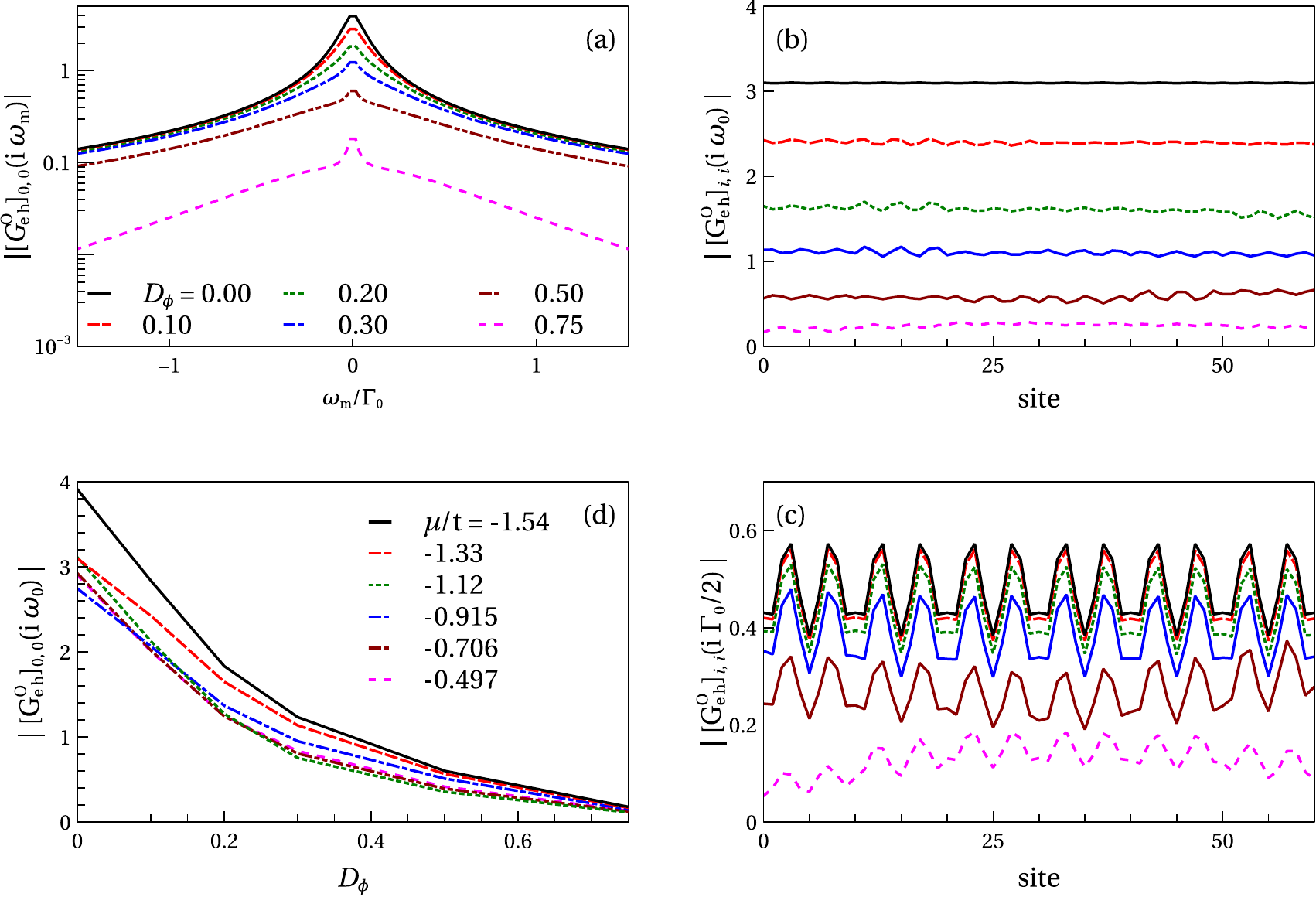}
\caption{\label{fig:GehOdd} Dependence of the magnitude of the on-site $[G^{\rm O}_{e h}]_{i, i}(\ii \, \omega_{m})$ on various physical parameters for a 40-unit chain of MZMs. (a): Phase disorder $D_{\phi}$ dependence as a function of Matsubara frequency $\omega_{m}$ for the pair amplitude under a MZM. (b,c): Phase disorder $D_{\phi}$ dependence as a function of site index $i$ for site locations covering 6 out of the 40 periodis in the chain for the lowest Matsubara frequency (b) and finite Matsubara frequency $\omega_{m} = {\rm \Gamma}_{0}/2$ (c). (d): Chemical potential $\mu$ dependence as a function of phase disorder $D_\phi$. In (a-c) $\mu/t = -1.54$, while the disorder averaging procedure and common parameters are given in the main text.}
\end{figure*}

We choose a ring of $M = 40$ MZMs, with $z = 10$ atomic sites between two neighboring MZMs. Since the folded bands alternate from electron-like to hole-like around momentum $p = 0$, we center the chemical potential to cut the third band ($l = 2$), as it is the first higher electron-like band, and choose a $k_{\rm F} = 0.2 \, \pi$.
We then also scan six equidistant values of the chemical potential, ranging from the third to the fifth band with the same $k_{\rm F}$, whose numerical values are given in Fig.~\ref{fig:GehOdd}(d). This assures that our results are not fine-tuned but can be interpreted as general.
As the band index increases, the bandwidth also increases (see e.g.~Fig.~\ref{fig:TBFold}). We therefore choose $\Gamma_{0}$ to be equal to $0.4$ of the smaller of the distances from the Fermi level to the band edge for the lowest scanned band, which turns out to be $\Gamma_{0} = 0.0815 \, t$. Since the effects of disorder in the coupling coefficients are investigated, we fix the energy splitting in each NW to a small finite value $\delta = 0.1 \, \Gamma_{0}$.

\begin{figure}[!ht]
\includegraphics[width=0.8\linewidth]{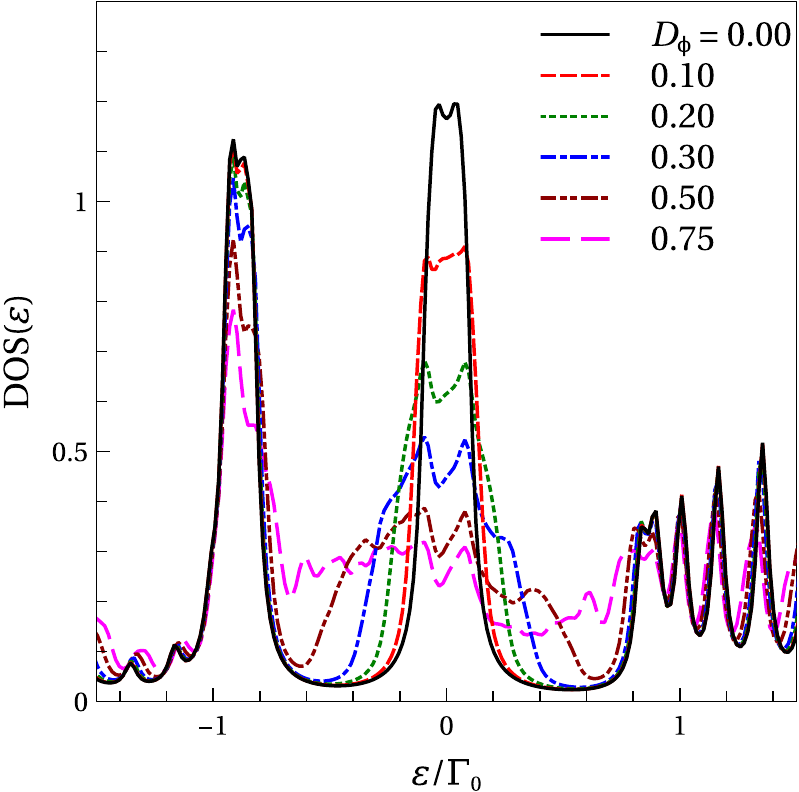}
\caption{\label{fig:DOS} DOS in the SPW and its dependence on the phase disorder $D_{\phi}$ for $\mu/t = -1.54$ for a 40-unit chain of MZM. Disorder averaging procedure and other parameters are given in the main text.}
\end{figure}

First of all, we verify that keeping the disorder real ($D_{\phi} = 0$) reproduces the robustness of the induced SC correlations, as also found in Ref.~\onlinecite{Lee16}, and that regardless of the value of $s$ (the sign bias of the disorder), and also for relatively large magnitude disorder, up to $D_{r} = 1$. Then, with the value of $\Gamma_{0}$ fixed, we generate $20$ different random disorder configurations according to the distribution described in the introduction of Sec.~\ref{sect1}, using the parameters $s = 1$,  $D_{r} = 0$, and with $D_{\phi}$ ranging through the values listed in Fig.~\ref{fig:GehOdd}. Here $D_{r}$  is a dimensionless measure of the disorder in the coupling given by the relative error of $|\Gamma|^{2}$ (see Appendix \ref{sec:ModelPDF}). The results of these calculations are summarized in Figs.~\ref{fig:GehOdd} and \ref{fig:DOS}. 

In Fig.~\ref{fig:GehOdd} we plot the magnitude of on-site, and thus dominating, odd-$\omega$  component of the sample-averaged anomalous Green's function $\left[G^{\rm O}_{e h}\right]_{i, i}(\ii \, \omega_{m})$ as a function of Matsubara frequency $\omega_{m}$, site location $i$, and $D_\phi$. The overall $\omega_{m}$-dependence exhibits a drastic suppression with phase-disorder strength $D_{\phi}$, as evident in the log-scale plot in Fig.~\ref{fig:GehOdd}(a). We stress that this behavior is thus very different for the effect of real disorder, where the pair amplitudes are robust.\cite{Lee16} 
The data in Fig.~\ref{fig:GehOdd}(a) are the odd-$\omega$ component directly under one of the MZMs. In Figs.~\ref{fig:GehOdd}(b) and (c), we plot the site dependence over $6$ out of the $40$ periods for both the lowest positive and a high frequency component. For low frequencies we see very little site variations. This can be understood from our analytical results for the single MZM, Eq.\,(\ref{GehSMZM}), where the decay length is $\sim 1/\omega_{m}$.\cite{RevModPhys.77.1321,Lee16}
At higher frequencies, here represented by Fig.~\ref{fig:GehOdd}(c) evaluated at half the bulk gap $\Gamma_{0}/2$, we find more variations with position. The anomalous Green's function $\left[ G^{\rm O}_{e h}\right]_{i ,i}(\ii \, \Gamma_{0}/2)$ actually has peaks (approximately) at a quarter distance from a MZM with dips both under a MZM, as well as at a mid-point between them.\footnote{According to the sampling theorem, the shortest wavelength that may be reproduced by a discrete set of values a distance $b/z$ apart is  $2b/z$, which, for $z = 10$, is equal to $1/5$ of the MZM distance.} 

Finally, in the last panel Fig.~\ref{fig:GehOdd}(d) we show how the low-frequency on-site component is suppressed with disorder for several different values of the chemical potential. Despite the doping-dependence of the DOS, we find a very similar monotonic suppression with phase-disorder strength, which shows that our results are not sensitive to the specifics of the model situation.
Beyond the dominant on-site odd-$\omega$ contribution to the anomalous Green's function plotted in Fig.~\ref{fig:GehOdd}, there are also sub-dominant non-local contributions, in analogy to the analytical results in Sec.~\ref{sec:Results2MZM}, which have both even-$\omega$, and odd-$\omega$ dependence and are peaked at different positions along the chain depending on the particular choice of model parameters. 
However, they are equivalently suppressed with the dominant on-site contribution as the phase disorder increases.

In Fig.~\ref{fig:DOS} we focus on the DOS in the SPW and plot the the site-averaged DOS for several values of the phase disorder $D_\phi$, for the same parameters as in Fig.~\ref{fig:GehOdd}(a-c). With increasing disorder we see that the near-zero-energy peaks (split off zero energy because a finite $\delta$) is diminishing but most notable is that the filling of the energy gap at finite energies.
A crucial difference here compared to conventional effects of pair-breaking disorder is that the gap gets filled by smearing of the near-zero-energy peaks and by states near the gap edge. Thus the in-gap states are naturally associated with the MZMs, and we interpret their shift in energy as the formation of non-local bound states corresponding to phase-matching conditions, similarly to the simplest example of two MZMs considered in Sec.~\ref{sec:Results2MZM}.

Another feature that is prominent in Fig.~\ref{fig:DOS} is the large peak that occurs near $-\Gamma_{0}$. This peak is more robust to the disorder than the energy gap. By carefully inspecting the site-resolved LDOS, we identify this as a contribution coming from sites that are at a quarter of the MZM distance from a particular MZM. Under an MZM, and at the midpoint between two MZMs, this peak has actually the smallest weight. Note here that we plot the electron DOS thus the DOS is not needed to be symmetric around zero energy. The missing DOS for creating a symmetric spectrum are hole-like quasiparticles.

\section{\label{concl}Conclusions}
In this work we have investigated the robustness of the superconducting pair correlations 
induced into a spin-polarized wire (SPW) from an array of Majorana zero modes (MZMs) when the couplings between these two systems are complex and acquire different phases. 
This corresponds to a  realistic situation, as complex couplings  generically appear for varying system parameters and disorder in the phases of the complex couplings is experimentally unavoidable.

First, we have demonstrated that, in general, the pair correlations induced into the SPW exhibit both even- and odd-frequency ($\omega$) dependence, which exponentially decay from the position of the MZMs. We have shown this effect for a single MZM, a pair of MZMs, and for an array of MZMs. Interestingly, we find that the phases of the complex couplings get transferred to the pair amplitudes.

Second, we have shown that the induced pair correlations, including the odd-$\omega$ component, suffer a considerable suppression due to phase-disorder averaging, exactly as a consequence of the transferred complex phases. This is in stark contrast to the effect of real disorder where the pair amplitudes remain robust.\cite{Lee16} We have found this strong disorder dependence both by analytically evaluating the $T$-matrix within the second Born approximation and performing numerical tight-binding calculations for an array of MZMs.

Third, we find that the suppression of the pair correlations in the SPW occurs concurrently with the filling of the energy gap by in-gap bound states appearing between spatially-separated MZMs. Carefully analyzing the situation of a pair of MZM coupled to a SPW, we have found bound states between the two MZMs that are highly sensitive to the relative complex phase between the two couplings. We can thus conclude that it is the complex phases and their disorder that is also causing the filling of the energy gap.

In summary, our work demonstrates that the conditions favorable for practical realization of a bulk 1D odd-$\omega$ superconductivity from MZMs requires full control of the system parameters. 

\section{Acknowledgements}
We thank J.~Klinovaja and P.~Burset for helpful discussions.  D.K.~acknowledges funding from the Knut and Alice Wallenberg foundation 
through the Wallenberg Academic Fellow grant of J.~Nilsson, J.C.~and A.M.B.-S.~acknowledge support from the Swedish Research Council (Vetenskapsr\aa det, Grant No.~2018-03488), the European Research Council (ERC) under the European Unions Horizon 2020 research and innovation programme (ERC- 2017-StG-757553), and the Knut and Alice Wallenberg Foundation through the Wallenberg Academy Fellows program.

\appendix
\section{\label{sec:ModelPDF}Distribution of coupling coefficients}
In this Appendix we give more details on the distribution used to generate the random coupling strengths. We treat the magnitude and phase of $\Gamma$ as independent. As for $\vert \Gamma \vert$, being a non-negative quantity, it is convenient to use the (natural) logarithm $\ln(\vert \Gamma \vert/\Gamma_{0})$, which spans the whole real line. We choose the distribution of the logarithm to be uniform $\mathcal{U}(f_{\mathrm{min}}, f_{\mathrm{max}})$, with the bounds $f_{\mathrm{min}}$ and $f_{\mathrm{max}}$ chosen such that:
\begin{subequations}
\label{eq:RealDisorderConsts}
\begin{eqnarray}
& \avg{ \vert \Gamma \vert^{2}} = \Gamma^{2}_{0}, \label{eq:RealDisorderAvg} \\
& \sqrt{\frac{\avg{\vert \Gamma \vert^{4}}}{(\avg{\vert \Gamma \vert^{2}})^{2}} - 1} = D_{r} \ge 0. \label{eq:RealDisorderRelUnc}
\end{eqnarray}
\end{subequations}
which leads to the the following expressions:
\begin{subequations}
\label{eq:RealDisroderParams}
\begin{eqnarray}
f_{\mathrm{min}} = -\frac{1}{2} \, \ln \left(\frac{\ee^{2 \, x} - 1}{2 \, x}\right), \label{eq:RealDisorderParam1} \\
f_{\mathrm{max}} = -\frac{1}{2} \, \ln \left(\frac{1 - \ee^{-2 \, x}}{2 \, x}\right), \label{eq:RealDisorderParam2} \\
x \, \coth(x) = 1 + D^{2}_{r}. \label{eq:RealDisorderParam3}
\end{eqnarray}
\end{subequations}
Here, $\Gamma_{0}$ is to be considered as a constant effective coupling that would reproduce the same $T$-matrix in \reqs{eq:Vertex} and \rref{eq:TmatrixDef}, $D_{r}$ is a dimensionless measure of the disorder in the coupling given by the relative error of $\vert \Gamma \vert^{2}$.
In the case of real couplings, the signs of $\Gamma$ do not enter \req{eq:Vertex}, and, thus, play no effect on the SPW. Nevertheless, we model a sign bias by a parameter $s$, $-1 \le s \le 1$, being the difference in the weight of positive and negative signs.

For creating a distribution of complex couplings $\Gamma = \vert \Gamma \vert \, \ee^{\ii \, \theta}$, we choose, strictly by convenience, that the phase follows a weighted uniform distribution around $0$ with weight $(1 + s)/2$, and symmetrically around $\pi$ and $-\pi$ with weight $(1 - s)/4$:
\begin{eqnarray}
& P( \theta ) = \frac{1 + s}{2} \, W\left(\frac{2 \, \theta}{\pi \, D_{\phi}} \right) + \nonumber \\
& +  \frac{1 - s}{4} \, \left\lbrace W\left[ \frac{4}{D_{\phi}} \, \left(\frac{\theta}{\pi} - 1 \right) + 1 \right] \right. \nonumber \\
& \left. + W\left[ \frac{4}{D_{\phi}} \, \left(\frac{\theta}{\pi} + 1\right) - 1 \right]\right\rbrace, \label{eq:DisorderPhaseDist}
\end{eqnarray}
where $W(x) = (1/2) \, \Theta(1 - \vert x \vert)$ is a window function from $-1$ to $1$ with unit weight. Here $D_{\phi}$ is the total fraction of the $2\pi$ phase interval that is covered. We use the distribution $P(\theta)$ to numerically model phase disorder in the couplings in Section \ref{sec:ResultsNumerics} of the main text.

\section{\label{sec:AppMZMBound}Bound states for two MZMs}
\begin{figure}[!h]
\includegraphics[width=0.95\linewidth]{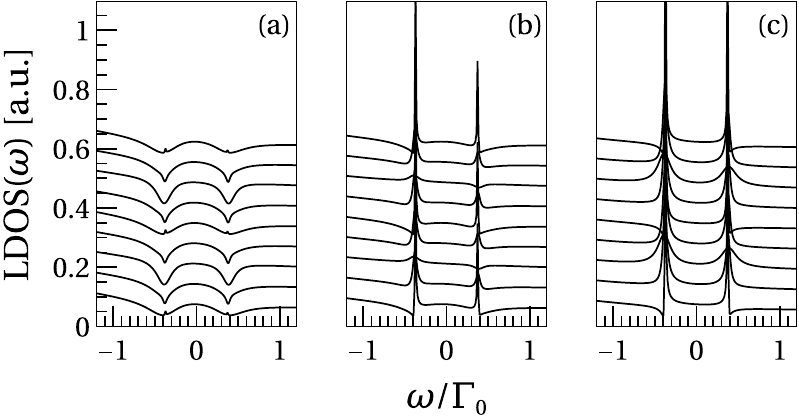}
\caption{\label{fig:2MZMLDOSb=0.5}Everything is the same as in Fig.~\ref{fig:2MZMLDOSb=1}, except $b = \lambda_{\rm F}/2$.}
\end{figure}

\begin{figure}[!h]
\includegraphics[width=0.95\linewidth]{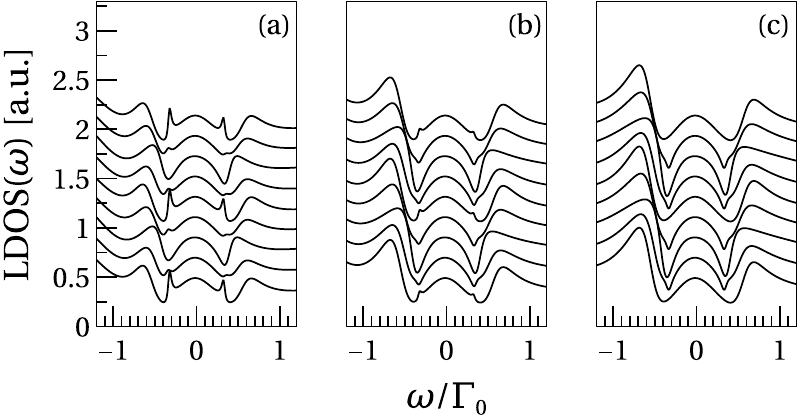}
\caption{\label{fig:2MZMLDOSb=2}Everything is the same as in Fig.~\ref{fig:2MZMLDOSb=1}, except $b = 2 \, \lambda_{\rm F}$.}
\end{figure}

In this Appendix we present complementary information for the delocalized bound states in the case of $2$ MZMs coupled to a SPW, originally discussed in Fig.~\ref{fig:2MZMLDOSb=1}. In Figs.~\ref{fig:2MZMLDOSb=0.5} and \ref{fig:2MZMLDOSb=2} we present the same spectrum as Fig.~\ref{fig:2MZMLDOSb=1}, but now for $b = 0.5 \, \lambda_{\rm F}$, and $b = 2 \, \lambda_{\rm F}$. Clearly, the relative position of the nodes and antinodes changes with this change of MZM-MZM distance. 
In Fig.~\ref{fig:2MZMLDOSb=0.5}, the nodes occur at $x = 0$ (and $x = b$), while there are peaks or antinodes both at $x = b/2$ and at $x = b/4$ (and $x = 3 b/4$), but clearly highest at $x= b/2$. This suggests that the wavelength of the standing wave, $\lambda_{s}$, is $\lambda_{s}/2 = b$, or $\lambda_{s} = 2 \, b$. Because $b = \lambda_{\rm F}/2$ in this case, $\lambda_{s} = \lambda_{\rm F}$. 
On the other hand, in Fig.~\ref{fig:2MZMLDOSb=2} it there are nodes on all panels $x = 0$, $x = b/4$, $x = b/2$ (and, by symmetry at $x = 3 b/4$ and $x = b$ as well). This implies $2 \, \lambda_{s} = b$, or $\lambda_{s} = b/2$. Because $b = 2 \, \lambda_{\rm F}$ in this case, $\lambda_{s} = \lambda_{\rm F}$. This shows how the periodicity is set by the Fermi wavelength as stated in the main text.

\section{\label{sec:AppDisAvg} Derivation of the disorder-averaged $T$-matrix}
In this Appendix, we derive a working expression for $\avg{\hat{\tilde{T}}}(p; \ii \, \omega_{m})$, the disordered-average of the matrix $T$ acting as an impurity $T$-matrix. The starting point is \req{eq:TmatFourierEq} in the main text, which acquires the shorthand form:
\begin{equation}
\label{eq:TmatrixEqFourierShort}
\left[ 1 - V \cdot g_{0} \right] \cdot T = V.
\end{equation}
Here, every symbol is a matrix in Nambu and momentum space, and matrix multiplication involves an integration over a dummy momentum variable ($\int_{-\pi}^{\pi} \frac{dp''}{2\pi}$), and the $1$ is an identity matrix in both Nambu and momentum space, with the Dirac delta in momentum space containing an extra factor of $2\pi$ ($1 \rightarrow 2\pi \, \delta(p - p'') \, \hat{1}$). With this, all the rules of matrix algebra are readily applicable to \req{eq:TmatrixEqFourierShort}.
In particular, \req{eq:TmatrixEqFourierShort} can be rewritten in an expanded form:
\[
T = V + V \cdot g_{0} \cdot T.
\]
Expressing both $T$ and $V$ as their expectation values plus a deviation from the expectation value we get:
\begin{eqnarray*}
& \avg{T} + \delta T = \avg{V} + \delta V + \\
& + \avg{V} \cdot g_{0} \cdot \avg{T} + \delta V \cdot g_{0} \avg{T} + \\
& + \avg{V} \cdot g_{0} \cdot \delta T + \delta V \cdot g_{0} \cdot \delta T.
\end{eqnarray*}
Then, taking the expectation value, and keeping in mind that $\avg{\delta V} = \avg{\delta T} = 0$ by definition, we arrive at the following equation for the expectation value:
\begin{equation}
\label{eq:TmatrixAv}
\avg{T} = \avg{V} + \avg{V} \cdot g_{0} \cdot \avg{T} + \avg{\delta V \cdot g_{0} \cdot \delta T}.
\end{equation}
Subtracting this equation from the full equation, we arrive at the equation for the deviations:
\begin{eqnarray}
\label{eq:TmatrixDev}
& \delta T = \delta V + \delta V \cdot g_{0} \avg{T} + \avg{V} \cdot g_{0} \cdot \delta T + \nonumber \\
& + \delta V \cdot g_{0} \cdot \delta T - \avg{\delta V \cdot g_{0} \cdot \delta T}.
\end{eqnarray}
We see here that to find $\avg{T}$ we need to evaluate the higher moment $\avg{\delta V \cdot g_{0} \cdot \delta T}$ in \req{eq:TmatrixAv}. To calculate that average, we reformulate \req{eq:TmatrixDev} in a more convenient form in terms of $g_{0} \cdot \delta T$ as:
\begin{subequations}
\label{eq:TmatrixDevNew}
\begin{eqnarray}
& g_{0} \cdot \delta T = g_{1} \cdot \left[ \delta V + \delta V \cdot g_{0} \cdot \avg{T} + \right. \nonumber \\
& \left. + \delta V \cdot g_{0} \cdot \delta T - \avg{\delta V \cdot g_{0} \cdot \delta T} \right], \label{eq:TmatrixDev2} \\
& g^{-1}_{1} = g^{-1}_{0} - \avg{V}. \label{eq:g1def}
\end{eqnarray}
\end{subequations}
The newly introduced matrix $g_{1}$ incorporates the averaged out effects of the MZM array and is diagonal in momentum space. Using \req{eq:g1def}, \req{eq:TmatrixAv} may be rewritten as:
\begin{equation}
\label{eq:TmatrixAv2}
g_{0} \cdot \avg{T} = g_{1} \cdot \left[ \avg{V} + \avg{\delta V \cdot g_{0} \cdot \delta T} \right].
\end{equation}
Now by multiplying \req{eq:TmatrixDev2} by $\delta V$ from the left and taking an average, we see that $\avg{\delta V \cdot g_{0} \cdot \delta T}$ is expressed in terms of the new averages $\avg{\delta V \cdot g_{1} \cdot \delta V}$, and $\avg{\left(\delta V \cdot g_{1}\right) \cdot \delta V \cdot g_{0} \cdot \delta T}$:
\begin{eqnarray*}
& \avg{\delta V \cdot g_{0} \cdot \delta T} = \avg{\delta V \cdot g_{1} \cdot \delta V} + \\
& + \avg{\delta V \cdot g_{1} \cdot \delta V} \cdot g_{0} \cdot \avg{T} + \\
& + \avg{\left(\delta V \cdot g_{1}\right) \cdot \delta V \cdot g_{0} \cdot \delta T}.
\end{eqnarray*}
Repeating the same procedure by multiplying \req{eq:TmatrixDev2} by $\left(\delta V \cdot g_{1} \right) \cdot \delta V$ and taking an average, we see that the average $\avg{\left(\delta V \cdot g_{1}\right) \cdot \delta V \cdot g_{0} \cdot \delta T}$ is expressed in terms of the new average $\avg{\left(\delta V \cdot g_{1}\right)^{2} \cdot \delta V \cdot g_{0} \cdot \delta T}$:
\begin{eqnarray*}
& \avg{\left(\delta V \cdot g_{1}\right) \cdot \delta V \cdot g_{0} \cdot \delta T} = \avg{\left(\delta V \cdot g_{1}\right)^{2} \cdot \delta V \cdot g_{0} \cdot \delta T} - \\
& - \avg{\delta V \cdot g_{1} \cdot \delta V} \cdot g_{1} \cdot \avg{\delta V \cdot g_{0} \cdot \delta T}.
\end{eqnarray*}
As a consequence it means that the following recursive relation for the averages $X_{n} = \avg{\left(\delta V \cdot g_{1}\right)^{2 n} \cdot \delta V \cdot g_{0} \cdot \delta T}$, involving the expectation values $\Pi^{\ast}_{n} = \avg{\left(\delta V \cdot g_{1}\right)^{2 n - 1} \cdot \delta V}$, holds:
\begin{eqnarray}
\label{eq:TmatrixDevRecur}
& X_{n} - X_{n + 1} = \nonumber \\
& = \Pi^{\ast}_{n + 1} + \Pi^{\ast}_{n + 1} \cdot \left( g_{0} \cdot \avg{T} - g_{1} \cdot X_{0} \right), \nonumber \\
& = \Pi^{\ast}_{n + 1} + \Pi^{\ast}_{n + 1} \cdot g_{1} \cdot \avg{V}, \ \ n \ge 0,
\end{eqnarray}
where we have also used \req{eq:TmatrixAv2}.
Summing from $n = 0$ to $\infty$ in \req{eq:TmatrixDevRecur}, and defining $\Pi^{\ast} = \sum_{n = 1}^{\infty} \Pi^{\ast}_{n}$, we get the following equation for $X_{0}$:
\begin{eqnarray*}
& X_{0} = \avg{\delta V \cdot g_{0} \cdot \delta T} \\
& = \Pi^{\ast}  + \Pi^{\ast} \cdot g_{1} \cdot \avg{V}.
\end{eqnarray*}
Plugging this equation into \req{eq:TmatrixAv2}, and keeping in mind that $g_{1} \cdot \avg{V}$ equals $g_{1} \cdot g^{-1}_{0} - 1$ according to \req{eq:g1def}, we have:
\begin{eqnarray*}
& g_{0} \cdot \avg{T} = g_{1} \cdot g^{-1}_{0} - 1 + g_{1} \cdot \Pi^{\ast} + \\
& + g_{1} \cdot \Pi^{\ast} \cdot g_{1} \cdot g^{-1}_{0} - g_{1} \cdot \Pi^{\ast},
\end{eqnarray*}
\begin{equation}
\label{eq:gavgdef}
\avg{g} = g_{1} + g_{1} \cdot \Pi^{\ast} \cdot g_{1} = g_{0} + g_{0} \cdot \avg{T} \cdot g_{0}.
\end{equation}
\req{eq:gavgdef} simultaneously defines a ``dirty'' discrete Green's function $\avg{g}$ for the SPW, incorporating the effects of any disorder in the matrices $V_{n}$ on top of an averaged-out homogeneous coupling $\avg{V}$, as well as an implicit equation for the ``dirty'' discrete $T$-matrix $\avg{T}$.

At this point it is customary to introduce the one-particle-irreducible self-energies:
\begin{equation}
\label{eq:1PIselfenergy}
\Pi = \sum_{n = 1}^{\infty} \avg{\left(\delta V \cdot g_{1}\right)^{2 n - 1} \cdot \delta V}_{\mathrm{1PI}},
\end{equation}
where with the Wick contractions we retain only the ones that cannot be disconnected by ``cutting'' a $g_{1}$ line. It is obvious that this self energy satisfies:
\begin{equation}
\label{eq:SelfEnergy}
\Pi^{\ast} = \Pi + \Pi \cdot g_{1} \cdot \Pi^{\ast},
\end{equation}
which implies a Dyson equation for $\avg{g}$:
\begin{equation}
\label{eq:gavgDyson}
\avg{g} = g_{1} + g_{1} \cdot \Pi \cdot \avg{g} \Leftrightarrow \avg{g}^{-1} = g^{-1}_{1} - \Pi.
\end{equation}
By using \reqs{eq:g1def} and \rref{eq:gavgDyson} we may write:
\begin{equation}
\label{eq:sigmadef}
\avg{g}^{-1} = g^{-1}_{0} - \Sigma, \ \Sigma = \avg{V} + \Pi,
\end{equation}
which defines the self-energy.
Then, using the second equality in \req{eq:gavgdef} for $\avg{T}$, after some algebra, we obtain:
\begin{equation}
\label{eq:TmatrixAvgDyson}
\avg{T}^{-1} = \Sigma^{-1} - g_{0}.
\end{equation}
If this equation is compared with an alternative form of \req{eq:TmatrixEqFourierShort}, namely $T^{-1} = V^{-1} - g_{0}$, we see that the role of $V^{-1}$ after averaging over disorder is played by $\Sigma^{-1}$, or equivalently, the inverse of the self-energy is the inverse of $V$.

\section{\label{sec:AppSelfEn}Evaluation of the $T$-matrix}
In this Appendix we evaluate the disorder-averaged $T$-matrix. The necessary steps for evaluating $\avg{T}$ are:
\begin{enumerate}
	\item{
		Calculate $\avg{V}$ from \req{eq:VmatrixFourier} and \req{eq:Vertex}. This already involves evaluating the first moment, so retaining only this expectation value amounts to the first Born approximation;
	}
	\item{
		Calculate $g_{1}$ from \req{eq:g1def} and from the obtained result for $\avg{V}$;
	}
	\item{
		Calculate $\Pi$ by truncating the sum in \req{eq:1PIselfenergy} and from the obtained result for $g_{1}$. Retaining only the first term involves calculating the second moment, so it amounts to the second Born approximation;
	}
	\item{
		Calculate $\Sigma_{1}$ to within the same approximation from \req{eq:sigmadef} and using $\Pi_{1}$ and $\avg{V}$;
	}
	\item{
		Finally, use \req{eq:TmatrixAvgDyson} to calculate $\avg{T_{1}}$ within the same approximation.
	}
\end{enumerate}
We now go ahead and follow each of the above steps, but also reinstate the explicit $p$ and $\omega_{m}$ dependence and the matrix structure in Nambu space.

The first moment $\avg{V}$ is :
\begin{eqnarray}
& \hat{\tilde{\avg{V}}}(p, p'; \ii \, \omega_{m}) = \avg{\hat{\tilde{V}}(p - p'; \ii \, \omega_{m})} \nonumber \\
& = \sum_{n} \ee^{-\ii \, (p - p') \, n} \, \avg{\hat{V}_{n}(\ii \, \omega_{m})} \nonumber \\
& = 2\pi \, \delta(p - p') \, \avg{\hat{V}}(\ii \, \omega_{m}), \label{eq:1Born}
\end{eqnarray}
with:
\begin{subequations}
\label{eq:FirstMoment}
\begin{eqnarray}
& \avg{\hat{V}}(\ii \, \omega_{m}) = \phi_{1}(\ii \, \omega_{m}) \, \left(\begin{array}{cc}
A_{1} & B_{1} \\
B^{\ast}_{1} & A_{1}
\end{array}\right), \label{eq:FirstMomentMatrix} \\
& \phi_{1}(\ii \, \omega_{m}) = \avg{\frac{-\ii \, \omega_{m}}{2 \, (\omega^{2}_{m} + \delta^{2})}}_{\delta}, \label{eq:DeltaAv1} \\
& A_{1} = \avg{\vert \Gamma \vert^{2}}_{\Gamma}, \label{eq:GammaAv1a} \\
& B_{1} =  \avg{\Gamma^{2}}_{\Gamma}, \label{eq:GammaAv2a}
\end{eqnarray}
\end{subequations}
where the subscript in the averages denotes over which disordered variable (the coupling coefficients $\Gamma$ or the energy splitting $\delta$ in the NW) the average is taken.

We notice that the electron-hole (and hole-electron) matrix elements are proportional to an average $B_{1}$ that is significantly suppressed under phase averaging because it carries a phase factor $\ee^{2 \ii \theta}$ as opposed to the average $A_{1}$ which is phase independent. Furthermore, if any off-diagonal matrix-element is still non-zero after averaging, it is odd in frequency because the average \req{eq:DeltaAv1} is odd in frequency.

Next, we evaluate $g_{1}$. Since both $g_{0}$ and $\avg{V}$ are proportional to $2\pi \, \delta(p - p')$, it follows that $\hat{g}_{1}(p, p'; \ii \, \omega_{m}) = \hat{g}_{1}(p; \ii \, \omega_{m}) \, 2\pi \, \delta(p - p')$, where:
\begin{subequations}
\label{eq:g1matrix}
\begin{eqnarray}
& \hat{g}_{1}(p; \ii \, \omega_{m}) = \begin{pmatrix}
 g_{1}^{ee}(p; \ii \, \omega_{m}) & g_{1}^{eh}(p; \ii \, \omega_{m}) \\
 g_{1}^{he}(p; \ii \, \omega_{m}) & g_{1}^{hh}(p; \ii \, \omega_{m})
\end{pmatrix}, \\
& g_{1}^{ee}(p; \ii \, \omega_{m}) = \frac{g_{e}(p; \ii \, \omega_{m}) \, \left( 1 - \phi_{1}(\ii \, \omega_{m}) \, A_{1} g_{h}(p; \ii \, \omega_{m}) \right)}{D_{1}(p; \ii \, \omega_{m})}, \\
& g_{1}^{eh}(p; \ii \, \omega_{m}) = \frac{g_{e}(p; \ii \, \omega_{m}) \, g_{h}(p; \ii \, \omega_{m}) \, \phi_{1}(\ii \, \omega_{m}) \, B_{1}}{D_{1}(p; \ii \, \omega_{m})}, \\
& g_{1}^{he}(p; \ii \, \omega_{m}) = \frac{g_{e}(p; \ii \, \omega_{m}) \, g_{h}(p; \ii \, \omega_{m}) \, \phi_{1}(\ii \, \omega_{m}) \, B^{\ast}_{1}}{D_{1}(p; \ii \, \omega_{m})}, \\
& g_{1}^{hh}(p; \ii \, \omega_{m}) = \frac{g_{h}(p; \ii \, \omega_{m}) \, \left( 1 - \phi_{1}(\ii \, \omega_{m}) \, A_{1} g_{e}(p; \ii \, \omega_{m}) \right)}{D_{1}(p; \ii \, \omega_{m})}, \\
& D_{1}(p; \ii \, \omega_{m}) = 1 - \phi_{1}(\ii \, \omega_{m}) \, A_{1} \times \nonumber \\
& \times \left( g_{e}(p; \ii \, \omega_{m}) + g_{h}(p; \ii \, \omega_{m}) \right) \nonumber \\
& + \phi^{2}_{1}(\ii \, \omega_{m}) \, \left( A^{2}_{1} - \vert B_{1} \vert^{2} \right) \, g_{e}(p; \ii \, \omega_{m}) \, g_{h}(p; \ii \, \omega_{m}).
\end{eqnarray}
\end{subequations}

\begin{widetext}
Then, in order to evaluate $\Pi_{1}$, we use an expectation value of the form:\begin{eqnarray}
& \avg{\delta \hat{V} \cdot \hat{X} \cdot \delta \hat{V}}(p, p'; \ii \, \omega_{m}) \nonumber \\
& = \avg{\hat{V} \cdot \hat{X} \cdot \hat{V}}(p, p'; \ii \, \omega_{m}) - \avg{\hat{V}} \cdot \hat{X} \cdot \avg{\hat{V}}(p, p'; \ii \, \omega_{m}) \nonumber \\
& = \int\int_{-\pi}^{\pi} \frac{dp_{1} \, dp_{2}}{(2\pi)^{2}} \, \left[ \avg{\hat{\tilde{V}}(p - p_{1}; \ii \, \omega_{m}) \cdot \hat{X}(p_{1}, p_{2}; \ii \, \omega_{m}) \cdot \hat{\tilde{V}}(p_{2} - p'; \ii \, \omega_{m})} \right. \nonumber \\
& \left. (2\pi)^{2} \, \delta(p - p_{1}) \, \delta(p_{2} - p') \, \avg{\hat{V}}(\ii \, \omega_{m}) \cdot \hat{X}(p_{1}, p_{2}; \ii \, \omega_{m}) \cdot \hat{V}(\ii \, \omega_{m}) \right] \nonumber \\
& = \int\int_{-\pi}^{\pi} \frac{dp_{1} \, dp_{2}}{(2\pi)^{2}} \sum_{n = -\infty}^{\infty} \ee^{-\ii \, (p - p_{1} + p_{2} - p') \, n} \times \nonumber \\
& \times \avg{\hat{V}_{n}(\ii \, \omega_{m}) \cdot \hat{X}(p_{1},p_{2}; \ii \, \omega_{m}) \cdot \hat{V}_{n}(\ii \, \omega_{m})} - \avg{\hat{V}} \cdot \hat{X}(p, p') \cdot \avg{\hat{V}} \nonumber \\
& = \int_{-\pi}^{\pi} \frac{d p''}{2\pi} \, \avg{\hat{V}(\ii \, \omega_{m}) \cdot \hat{X}(p'', p'' - p + p'; \ii \, \omega_{m}) \cdot \hat{V}(\ii \, \omega_{m})} \nonumber \\
& - \avg{\hat{V}}(\ii \, \omega_{m}) \cdot \hat{X}(p, p'; \ii \, \omega_{m}) \cdot \avg{\hat{V}}(\ii \, \omega_{m}),
\end{eqnarray}
where $\hat{X}(p, p'; \ii \, \omega_{m}) = \hat{X}(p + 2\pi, p'; \ii \, \omega_{m}) = \hat{X}(p, p'+2\pi; \ii \, \omega_{m})$ is a general matrix in Nambu and momentum space.

If the matrix is diagonal in momentum, such as the case for the matrix $\hat{g}_{1}$ the expression simplifies. Thus, the second Born approximation for the disorder-averaged self-energy is $\hat{\Pi}_{1}(p, p'; \ii \, \omega_{m}) = \hat{\Pi}_{1}(p; \ii \, \omega_{m}) \, 2\pi \, \delta(p - p')$ with:
\begin{eqnarray}
\label{eq:SelfEnFirstBorn1}
\hat{\Pi}_{1}(p; \ii \, \omega_{m}) & = \int_{-\pi}^{\pi}\frac{p''}{2\pi} \, \avg{\hat{V}(\ii \, \omega_{m}) \cdot \hat{g}_{1}(p''; \ii \, \omega_{m}) \cdot \hat{V}(\ii \, \omega_{m})} \nonumber \\
 & - \avg{\hat{V}}(\ii \, \omega_{m}) \cdot \hat{g}_{1}(p; \ii \, \omega_{m})\cdot \avg{\hat{V}}(\ii \, \omega_{m}).
\end{eqnarray}

The first term in \req{eq:SelfEnFirstBorn1} has the form:
\begin{eqnarray}
\label{eq:SecondMomentMatrix}
&\int_{-\pi}^{\pi} \frac{d p''}{2\pi} \,\avg{\hat{V}(\ii \, \omega_{m}) \cdot \hat{g}_{1}(p''; \ii \, \omega_{m}) \cdot \hat{V}(\ii \, \omega_{m})} = \phi_{2}(\ii \, \omega_{m}) \, \left(\begin{array}{cc}
Z_{e e}(\ii \, \omega_{m}) & Z_{e h}(\ii \, \omega_{m}) \\
Z_{h e}(\ii \, \omega_{m}) & Z_{h h}(\ii \, \omega_{m})
\end{array}\right),
\end{eqnarray}
where
\begin{subequations}
\label{eq:SecondMomentMatrixAvgs}
\begin{eqnarray}
& Z_{e e}(\ii \, \omega_{m}) = Z_{h h}(\ii \, \omega_{m}) = A_{2} \, \left(\avg{g_{1}^{e e}}(\ii \, \omega_{m}) + \avg{g_{1}^{h h}}(\ii \, \omega_{m})\right) \nonumber \\
& + B_{2} \, \avg{g_{1}^{h e}}(\ii \, \omega_{m}) + B^{\ast}_{2} \, \avg{g_{1}^{e h}}(\ii \, \omega_{m}), \label{eq:SecondMomentEE} \\
& Z_{e h} = A_{2} \, \avg{g_{1}^{eh}}(\ii \, \omega_{m}) + B_{2} \, \left(\avg{g_{1}^{ee}}(\ii \, \omega_{m}) + \avg{g_{1}^{h h}}(\ii \, \omega_{m})\right) + C_{2} \, \avg{g_{1}^{h e}}(\ii \, \omega_{m}), \label{eq:SecondMomentEH} \\
& Z_{h e} = A_{2} \, \avg{g_{1}^{he}}(\ii \, \omega_{m}) + B^{\ast}_{2} \, \left(\avg{g_{1}^{ee}}(\ii \, \omega_{m}) + \avg{g_{1}^{hh}}(\ii \, \omega_{m}) \right) + C^{\ast}_{2} \, \avg{g_{1}^{e h}}(\ii \, \omega_{m}), \label{eq:SecondMomentHE} \\
& \phi_{2}(\ii \, \omega_{m}) = -\avg{\frac{\omega^{2}_{m}}{4 \, ( \omega^{2}_{m} + \delta^{2})^{2}}}_{\delta} = \phi_{2}(-\ii \, \omega_{m}), \label{eq:Phi2} \\
& A_{2} = \avg{\vert \Gamma \vert^{4}}_{\Gamma}, \label{eq:A2} \\
& B_{2} = \avg{\vert \Gamma \vert^{2} \, \Gamma^{2}}_{\Gamma}, \label{eq:B2} \\
& C_{2} = \avg{\Gamma^{4}}_{\Gamma}, \label{eq:C2}
\end{eqnarray}
\end{subequations}
where we used the shorthand notation:
\[
\avg{\hat{g}_{1}}(\ii \, \omega_{m}) = \int_{-\pi}^{\pi} \frac{dp''}{2 \pi} \, \hat{g}_{1}(p''; \ii \, \omega_{m}).
\]
The second term in \req{eq:SelfEnFirstBorn1}, using \reqs{eq:FirstMoment}, \rref{eq:g1matrix} is:
\begin{eqnarray}
& \avg{\hat{V}}(\ii \, \omega_{m}) \cdot \hat{g}_{1}(p; \ii \, \omega_{m}) \cdot \avg{\hat{V}}(\ii \, \omega_{m}) = \nonumber \\
& \frac{\phi^{2}_{1}(\ii \, \omega_{m})}{D_{1}(p; \ii \, \omega_{m})} \, \left(\begin{array}{cc}
P^{ee}(p; \ii \, \omega_{m}) & P^{eh}(p; \ii \, \omega_{m}) \\
P^{he}(p; \ii \, \omega_{m}) & P^{hh}(p; \ii \, \omega_{m})
\end{array}\right), \label{eq:SecondMomentMatrix2}
\end{eqnarray}
where
\begin{subequations}
\label{eq:SecondMomentMatrix2Avgs}
\begin{eqnarray}
& P^{ee}(p; \ii \, \omega_{m}) = A^{2}_{1} \, g_{e}(p; \ii \, \omega_{m}) + \vert B_{1} \vert^{2} \, g_{h}(p; \ii \, \omega_{m}) - \phi_{1}(\ii \, \omega_{m}) \, A_{1} \, (A^{2}_{1} - \vert B_{1} \vert^{2}) \, g_{e}(p; \ii \, \omega_{m}) \, g_{h}(p; \ii \, \omega_{m}), \label{eq:SecondMoment2EE} \\
& P^{eh}(p; \ii \, \omega_{m}) = B_{1} \, \left[ A_{1} \, \left(g_{1}(p; \ii \, \omega_{m}) + g_{h}(p; \ii \, \omega_{m}) \right) - \phi_{1}(\ii \, \omega_{m}) \, (A^{2}_{1} - \vert B_{1} \vert^{2}) \, g_{e}(p; \ii \, \omega_{m}) \, g_{h}(p; \ii \, \omega_{m}) \right], \label{eq:SecondMoment2EH} \\
& P^{he}(p; \ii \, \omega_{m}) = B^{\ast}_{1} \, \left[ A_{1} \, \left(g_{1}(p; \ii \, \omega_{m}) + g_{h}(p; \ii \, \omega_{m}) \right) - \phi_{1}(\ii \, \omega_{m}) \, (A^{2}_{1} - \vert B_{1} \vert^{2}) \, g_{e}(p; \ii \, \omega_{m}) \, g_{h}(p; \ii \, \omega_{m}) \right], \label{eq:SecondMoment2HE} \\
& P^{hh}(p; \ii \, \omega_{m}) = A^{2}_{1} \, g_{h}(p; \ii \, \omega_{m}) + \vert B_{1} \vert^{2} \, g_{e}(p; \ii \, \omega_{m}) - \phi_{1}(\ii \, \omega_{m}) \, A_{1} \, (A^{2}_{1} - \vert B_{1} \vert^{2}) \, g_{e}(p; \ii \, \omega_{m}) \, g_{h}(p; \ii \, \omega_{m}). \label{eq:SecondMoment2HH}
\end{eqnarray}
\end{subequations}
All the terms in \req{eq:sigmadef} are in given by \reqs{eq:1Born}, \rref{eq:SelfEnFirstBorn1}, \rref{eq:SecondMomentMatrix}, and \rref{eq:SecondMomentMatrix2}. After some algebraic manipulation, we can show that \req{eq:TmatrixAvgDyson} has the solution:
\begin{eqnarray}
& \avg{\hat{T}_{1}}(p; \ii \, \omega_{m}) = \frac{1}{D_{T}(p; \ii \, \omega_{m})} \, \left(\begin{array}{cc}
N^{ee}_{T1}(p; \ii \, \omega_{m}) & N^{eh}_{T1}(p; \ii \, \omega_{m}) \\
N^{he}_{T1}(p; \ii \, \omega_{m}) & N^{hh}_{T1}(p; \ii \, \omega_{m})
\end{array}
\right), \label{eq:TmatExpr1} \\
\end{eqnarray}
where
\begin{subequations}
\label{eq:TmatExprs}
\begin{eqnarray}
& D_{T}(p; \ii \, \omega_{m}) = 1 - g_{e}(p; \ii \, \omega_{m}) \, \Sigma^{ee}_{1}(p; \ii \, \omega_{m}) - g_{h}(p; \ii \, \omega_{m}) \, \Sigma^{hh}_{1}(p; \ii \, \omega_{m}) \nonumber \\
& + g_{e}(p; \ii \, \omega_{m}) \, g_{h}(p; \ii \, \omega_{m}) \, {\rm det}(\hat{\Sigma}_{1})(p; \ii \, \omega_{m}), \label{eq:TmatDenom} \\
& N^{ee}_{T1}(p; \ii \, \omega_{m}) = \Sigma^{ee}_{1}(p; \ii \, \omega_{m}) - g_{h}(p; \ii \, \omega_{m}) \, {\rm det}(\hat{\Sigma}_{1})(p; \ii \, \omega_{m}), \label{eq:TmatNumEE}\\
& N^{eh}_{T1}(p; \ii \, \omega_{m}) = \Sigma^{eh}_{1}(p; \ii \, \omega_{m}), \  N^{he}_{T1}(p; \ii \, \omega_{m}) = \Sigma^{he}_{1}(p; \ii \, \omega_{m}), \label{eq:TmatNumOffDiag} \\
& N^{hh}_{T1}(p; \ii \, \omega_{m}) = \Sigma^{hh}_{1}(p; \ii \, \omega_{m}) - g_{h}(p; \ii \, \omega_{m}) \, {\rm det}(\hat{\Sigma}_{1})(p; \ii \, \omega_{m}),  \label{eq:TmatNumHH} \\
& {\rm det}(\hat{\Sigma}_{1})(p; \ii \, \omega_{m}) = \Sigma^{ee}_{1}(p; \ii \, \omega_{m}) \, \Sigma^{hh}_{1}(p; \ii \, \omega_{m}) - \Sigma^{eh}_{1}(p; \ii \, \omega_{m}) \, \Sigma^{he}_{1}(p; \ii \, \omega_{m}). \label{eq:detSigma}
\end{eqnarray}
\end{subequations}
\end{widetext}
Finally, to get \req{arrayanalyticEq}, we use the expression for $N^{eh}_{T1}$ from \req{eq:TmatNumOffDiag}, which tells us that we need the electron-hole element of  \req{eq:FirstMomentMatrix}, as well as \reqs{eq:SecondMomentEH} and \rref{eq:SecondMoment2EH}.

\bibliography{biblio}

\begin{thebibliography}{48}%
\makeatletter
\providecommand \@ifxundefined [1]{%
 \@ifx{#1\undefined}
}%
\providecommand \@ifnum [1]{%
 \ifnum #1\expandafter \@firstoftwo
 \else \expandafter \@secondoftwo
 \fi
}%
\providecommand \@ifx [1]{%
 \ifx #1\expandafter \@firstoftwo
 \else \expandafter \@secondoftwo
 \fi
}%
\providecommand \natexlab [1]{#1}%
\providecommand \enquote  [1]{``#1''}%
\providecommand \bibnamefont  [1]{#1}%
\providecommand \bibfnamefont [1]{#1}%
\providecommand \citenamefont [1]{#1}%
\providecommand \href@noop [0]{\@secondoftwo}%
\providecommand \href [0]{\begingroup \@sanitize@url \@href}%
\providecommand \@href[1]{\@@startlink{#1}\@@href}%
\providecommand \@@href[1]{\endgroup#1\@@endlink}%
\providecommand \@sanitize@url [0]{\catcode `\\12\catcode `\$12\catcode
  `\&12\catcode `\#12\catcode `\^12\catcode `\_12\catcode `\%12\relax}%
\providecommand \@@startlink[1]{}%
\providecommand \@@endlink[0]{}%
\providecommand \url  [0]{\begingroup\@sanitize@url \@url }%
\providecommand \@url [1]{\endgroup\@href {#1}{\urlprefix }}%
\providecommand \urlprefix  [0]{URL }%
\providecommand \Eprint [0]{\href }%
\providecommand \doibase [0]{http://dx.doi.org/}%
\providecommand \selectlanguage [0]{\@gobble}%
\providecommand \bibinfo  [0]{\@secondoftwo}%
\providecommand \bibfield  [0]{\@secondoftwo}%
\providecommand \translation [1]{[#1]}%
\providecommand \BibitemOpen [0]{}%
\providecommand \bibitemStop [0]{}%
\providecommand \bibitemNoStop [0]{.\EOS\space}%
\providecommand \EOS [0]{\spacefactor3000\relax}%
\providecommand \BibitemShut  [1]{\csname bibitem#1\endcsname}%
\let\auto@bib@innerbib\@empty
\bibitem [{\citenamefont {Berezinskii}(1974)}]{bere74}%
  \BibitemOpen
  \bibfield  {author} {\bibinfo {author} {\bibfnamefont {V.~L.}\ \bibnamefont
  {Berezinskii}},\ }\href
  {http://www.jetpletters.ac.ru/ps/1792/article_27363.shtml} {\bibfield
  {journal} {\bibinfo  {journal} {JETP Lett.}\ }\textbf {\bibinfo {volume}
  {20}},\ \bibinfo {pages} {287} (\bibinfo {year} {1974})}\BibitemShut
  {NoStop}%
\bibitem [{\citenamefont {Kirkpatrick}\ and\ \citenamefont
  {Belitz}(1991)}]{PhysRevLett.66.1533}%
  \BibitemOpen
  \bibfield  {author} {\bibinfo {author} {\bibfnamefont {T.~R.}\ \bibnamefont
  {Kirkpatrick}}\ and\ \bibinfo {author} {\bibfnamefont {D.}~\bibnamefont
  {Belitz}},\ }\href {\doibase 10.1103/PhysRevLett.66.1533} {\bibfield
  {journal} {\bibinfo  {journal} {Phys. Rev. Lett.}\ }\textbf {\bibinfo
  {volume} {66}},\ \bibinfo {pages} {1533} (\bibinfo {year}
  {1991})}\BibitemShut {NoStop}%
\bibitem [{\citenamefont {Balatsky}\ and\ \citenamefont
  {Abrahams}(1992)}]{PhysRevB.45.13125}%
  \BibitemOpen
  \bibfield  {author} {\bibinfo {author} {\bibfnamefont {A.}~\bibnamefont
  {Balatsky}}\ and\ \bibinfo {author} {\bibfnamefont {E.}~\bibnamefont
  {Abrahams}},\ }\href {\doibase 10.1103/PhysRevB.45.13125} {\bibfield
  {journal} {\bibinfo  {journal} {Phys. Rev. B}\ }\textbf {\bibinfo {volume}
  {45}},\ \bibinfo {pages} {13125} (\bibinfo {year} {1992})}\BibitemShut
  {NoStop}%
\bibitem [{\citenamefont {Linder}\ and\ \citenamefont
  {Balatsky}(2017)}]{Balatsky2017}%
  \BibitemOpen
  \bibfield  {author} {\bibinfo {author} {\bibfnamefont {J.}~\bibnamefont
  {Linder}}\ and\ \bibinfo {author} {\bibfnamefont {A.~V.}\ \bibnamefont
  {Balatsky}},\ }\href@noop {} {\bibfield  {journal} {\bibinfo  {journal}
  {arXiv:1709.03986}\ } (\bibinfo {year} {2017})}\BibitemShut {NoStop}%
\bibitem [{\citenamefont {Bergeret}\ \emph {et~al.}(2001)\citenamefont
  {Bergeret}, \citenamefont {Volkov},\ and\ \citenamefont
  {Efetov}}]{PhysRevLett.86.4096}%
  \BibitemOpen
  \bibfield  {author} {\bibinfo {author} {\bibfnamefont {F.~S.}\ \bibnamefont
  {Bergeret}}, \bibinfo {author} {\bibfnamefont {A.~F.}\ \bibnamefont
  {Volkov}}, \ and\ \bibinfo {author} {\bibfnamefont {K.~B.}\ \bibnamefont
  {Efetov}},\ }\href {\doibase 10.1103/PhysRevLett.86.4096} {\bibfield
  {journal} {\bibinfo  {journal} {Phys. Rev. Lett.}\ }\textbf {\bibinfo
  {volume} {86}},\ \bibinfo {pages} {4096} (\bibinfo {year}
  {2001})}\BibitemShut {NoStop}%
\bibitem [{\citenamefont {{Kadigrobov, A.}}\ \emph {et~al.}(2001)\citenamefont
  {{Kadigrobov, A.}}, \citenamefont {{Shekhter, R. I.}},\ and\ \citenamefont
  {{Jonson, M.}}}]{Kadigrobov01}%
  \BibitemOpen
  \bibfield  {author} {\bibinfo {author} {\bibnamefont {{Kadigrobov, A.}}},
  \bibinfo {author} {\bibnamefont {{Shekhter, R. I.}}}, \ and\ \bibinfo
  {author} {\bibnamefont {{Jonson, M.}}},\ }\href {\doibase
  10.1209/epl/i2001-00107-2} {\bibfield  {journal} {\bibinfo  {journal}
  {Europhys. Lett.}\ }\textbf {\bibinfo {volume} {54}},\ \bibinfo {pages} {394}
  (\bibinfo {year} {2001})}\BibitemShut {NoStop}%
\bibitem [{\citenamefont {Hashimoto}(2001)}]{PhysRevB.64.132507}%
  \BibitemOpen
  \bibfield  {author} {\bibinfo {author} {\bibfnamefont {K.}~\bibnamefont
  {Hashimoto}},\ }\href {\doibase 10.1103/PhysRevB.64.132507} {\bibfield
  {journal} {\bibinfo  {journal} {Phys. Rev. B}\ }\textbf {\bibinfo {volume}
  {64}},\ \bibinfo {pages} {132507} (\bibinfo {year} {2001})}\BibitemShut
  {NoStop}%
\bibitem [{\citenamefont {Di~Bernardo}\ \emph {et~al.}(2015)\citenamefont
  {Di~Bernardo}, \citenamefont {Salman}, \citenamefont {Wang}, \citenamefont
  {Amado}, \citenamefont {Egilmez}, \citenamefont {Flokstra}, \citenamefont
  {Suter}, \citenamefont {Lee}, \citenamefont {Zhao}, \citenamefont {Prokscha},
  \citenamefont {Morenzoni}, \citenamefont {Blamire}, \citenamefont {Linder},\
  and\ \citenamefont {Robinson}}]{PhysRevX.5.041021}%
  \BibitemOpen
  \bibfield  {author} {\bibinfo {author} {\bibfnamefont {A.}~\bibnamefont
  {Di~Bernardo}}, \bibinfo {author} {\bibfnamefont {Z.}~\bibnamefont {Salman}},
  \bibinfo {author} {\bibfnamefont {X.~L.}\ \bibnamefont {Wang}}, \bibinfo
  {author} {\bibfnamefont {M.}~\bibnamefont {Amado}}, \bibinfo {author}
  {\bibfnamefont {M.}~\bibnamefont {Egilmez}}, \bibinfo {author} {\bibfnamefont
  {M.~G.}\ \bibnamefont {Flokstra}}, \bibinfo {author} {\bibfnamefont
  {A.}~\bibnamefont {Suter}}, \bibinfo {author} {\bibfnamefont {S.~L.}\
  \bibnamefont {Lee}}, \bibinfo {author} {\bibfnamefont {J.~H.}\ \bibnamefont
  {Zhao}}, \bibinfo {author} {\bibfnamefont {T.}~\bibnamefont {Prokscha}},
  \bibinfo {author} {\bibfnamefont {E.}~\bibnamefont {Morenzoni}}, \bibinfo
  {author} {\bibfnamefont {M.~G.}\ \bibnamefont {Blamire}}, \bibinfo {author}
  {\bibfnamefont {J.}~\bibnamefont {Linder}}, \ and\ \bibinfo {author}
  {\bibfnamefont {J.~W.~A.}\ \bibnamefont {Robinson}},\ }\href {\doibase
  10.1103/PhysRevX.5.041021} {\bibfield  {journal} {\bibinfo  {journal} {Phys.
  Rev. X}\ }\textbf {\bibinfo {volume} {5}},\ \bibinfo {pages} {041021}
  (\bibinfo {year} {2015})}\BibitemShut {NoStop}%
\bibitem [{\citenamefont {Asano}\ and\ \citenamefont
  {Tanaka}(2013)}]{PhysRevB.87.104513}%
  \BibitemOpen
  \bibfield  {author} {\bibinfo {author} {\bibfnamefont {Y.}~\bibnamefont
  {Asano}}\ and\ \bibinfo {author} {\bibfnamefont {Y.}~\bibnamefont {Tanaka}},\
  }\href {\doibase 10.1103/PhysRevB.87.104513} {\bibfield  {journal} {\bibinfo
  {journal} {Phys. Rev. B}\ }\textbf {\bibinfo {volume} {87}},\ \bibinfo
  {pages} {104513} (\bibinfo {year} {2013})}\BibitemShut {NoStop}%
\bibitem [{\citenamefont {Cayao}\ \emph {et~al.}(2019)\citenamefont {Cayao},
  \citenamefont {Triola},\ and\ \citenamefont {Black-Schaffer}}]{cayao2019odd}%
  \BibitemOpen
  \bibfield  {author} {\bibinfo {author} {\bibfnamefont {J.}~\bibnamefont
  {Cayao}}, \bibinfo {author} {\bibfnamefont {C.}~\bibnamefont {Triola}}, \
  and\ \bibinfo {author} {\bibfnamefont {A.~M.}\ \bibnamefont
  {Black-Schaffer}},\ }\href@noop {} {\bibfield  {journal} {\bibinfo  {journal}
  {arXiv:1908.05466}\ } (\bibinfo {year} {2019})}\BibitemShut {NoStop}%
\bibitem [{\citenamefont {Kitaev}(2001)}]{kitaev}%
  \BibitemOpen
  \bibfield  {author} {\bibinfo {author} {\bibfnamefont {A.~Y.}\ \bibnamefont
  {Kitaev}},\ }\href@noop {} {\bibfield  {journal} {\bibinfo  {journal} {Phys.
  Usp.}\ }\textbf {\bibinfo {volume} {44}},\ \bibinfo {pages} {131} (\bibinfo
  {year} {2001})}\BibitemShut {NoStop}%
\bibitem [{\citenamefont {Elliott}\ and\ \citenamefont
  {Franz}(2015)}]{Franz2015}%
  \BibitemOpen
  \bibfield  {author} {\bibinfo {author} {\bibfnamefont {S.~R.}\ \bibnamefont
  {Elliott}}\ and\ \bibinfo {author} {\bibfnamefont {M.}~\bibnamefont
  {Franz}},\ }\href {\doibase 10.1103/RevModPhys.87.137} {\bibfield  {journal}
  {\bibinfo  {journal} {Rev. Mod. Phys.}\ }\textbf {\bibinfo {volume} {87}},\
  \bibinfo {pages} {137} (\bibinfo {year} {2015})}\BibitemShut {NoStop}%
\bibitem [{\citenamefont {Nayak}\ \emph {et~al.}(2008)\citenamefont {Nayak},
  \citenamefont {Simon}, \citenamefont {Stern}, \citenamefont {Freedman},\ and\
  \citenamefont {Das~Sarma}}]{RevModPhys.80.1083}%
  \BibitemOpen
  \bibfield  {author} {\bibinfo {author} {\bibfnamefont {C.}~\bibnamefont
  {Nayak}}, \bibinfo {author} {\bibfnamefont {S.~H.}\ \bibnamefont {Simon}},
  \bibinfo {author} {\bibfnamefont {A.}~\bibnamefont {Stern}}, \bibinfo
  {author} {\bibfnamefont {M.}~\bibnamefont {Freedman}}, \ and\ \bibinfo
  {author} {\bibfnamefont {S.}~\bibnamefont {Das~Sarma}},\ }\href@noop {}
  {\bibfield  {journal} {\bibinfo  {journal} {Rev. Mod. Phys.}\ }\textbf
  {\bibinfo {volume} {80}},\ \bibinfo {pages} {1083} (\bibinfo {year}
  {2008})}\BibitemShut {NoStop}%
\bibitem [{\citenamefont {Sarma}\ \emph {et~al.}(2015)\citenamefont {Sarma},
  \citenamefont {Freedman},\ and\ \citenamefont {Nayak}}]{Sarma:16}%
  \BibitemOpen
  \bibfield  {author} {\bibinfo {author} {\bibfnamefont {S.~D.}\ \bibnamefont
  {Sarma}}, \bibinfo {author} {\bibfnamefont {M.}~\bibnamefont {Freedman}}, \
  and\ \bibinfo {author} {\bibfnamefont {C.}~\bibnamefont {Nayak}},\
  }\href@noop {} {\bibfield  {journal} {\bibinfo  {journal} {Npj Quantum
  Information}\ }\textbf {\bibinfo {volume} {1}},\ \bibinfo {pages} {15001}
  (\bibinfo {year} {2015})}\BibitemShut {NoStop}%
\bibitem [{\citenamefont {Huang}\ \emph {et~al.}(2015)\citenamefont {Huang},
  \citenamefont {W\"olfle},\ and\ \citenamefont
  {Balatsky}}]{PhysRevB.92.121404}%
  \BibitemOpen
  \bibfield  {author} {\bibinfo {author} {\bibfnamefont {Z.}~\bibnamefont
  {Huang}}, \bibinfo {author} {\bibfnamefont {P.}~\bibnamefont {W\"olfle}}, \
  and\ \bibinfo {author} {\bibfnamefont {A.~V.}\ \bibnamefont {Balatsky}},\
  }\href {\doibase 10.1103/PhysRevB.92.121404} {\bibfield  {journal} {\bibinfo
  {journal} {Phys. Rev. B}\ }\textbf {\bibinfo {volume} {92}},\ \bibinfo
  {pages} {121404} (\bibinfo {year} {2015})}\BibitemShut {NoStop}%
\bibitem [{\citenamefont {Kashuba}\ \emph {et~al.}(2017)\citenamefont
  {Kashuba}, \citenamefont {Sothmann}, \citenamefont {Burset},\ and\
  \citenamefont {Trauzettel}}]{PhysRevB.95.174516}%
  \BibitemOpen
  \bibfield  {author} {\bibinfo {author} {\bibfnamefont {O.}~\bibnamefont
  {Kashuba}}, \bibinfo {author} {\bibfnamefont {B.}~\bibnamefont {Sothmann}},
  \bibinfo {author} {\bibfnamefont {P.}~\bibnamefont {Burset}}, \ and\ \bibinfo
  {author} {\bibfnamefont {B.}~\bibnamefont {Trauzettel}},\ }\href {\doibase
  10.1103/PhysRevB.95.174516} {\bibfield  {journal} {\bibinfo  {journal} {Phys.
  Rev. B}\ }\textbf {\bibinfo {volume} {95}},\ \bibinfo {pages} {174516}
  (\bibinfo {year} {2017})}\BibitemShut {NoStop}%
\bibitem [{\citenamefont {Alicea}(2010)}]{PhysRevB.81.125318}%
  \BibitemOpen
  \bibfield  {author} {\bibinfo {author} {\bibfnamefont {J.}~\bibnamefont
  {Alicea}},\ }\href {\doibase 10.1103/PhysRevB.81.125318} {\bibfield
  {journal} {\bibinfo  {journal} {Phys. Rev. B}\ }\textbf {\bibinfo {volume}
  {81}},\ \bibinfo {pages} {125318} (\bibinfo {year} {2010})}\BibitemShut
  {NoStop}%
\bibitem [{\citenamefont {Lutchyn}\ \emph {et~al.}(2010)\citenamefont
  {Lutchyn}, \citenamefont {Sau},\ and\ \citenamefont
  {Das~Sarma}}]{PhysRevLett.105.077001}%
  \BibitemOpen
  \bibfield  {author} {\bibinfo {author} {\bibfnamefont {R.~M.}\ \bibnamefont
  {Lutchyn}}, \bibinfo {author} {\bibfnamefont {J.~D.}\ \bibnamefont {Sau}}, \
  and\ \bibinfo {author} {\bibfnamefont {S.}~\bibnamefont {Das~Sarma}},\ }\href
  {\doibase 10.1103/PhysRevLett.105.077001} {\bibfield  {journal} {\bibinfo
  {journal} {Phys. Rev. Lett.}\ }\textbf {\bibinfo {volume} {105}},\ \bibinfo
  {pages} {077001} (\bibinfo {year} {2010})}\BibitemShut {NoStop}%
\bibitem [{\citenamefont {Oreg}\ \emph {et~al.}(2010)\citenamefont {Oreg},
  \citenamefont {Refael},\ and\ \citenamefont {von
  Oppen}}]{PhysRevLett.105.177002}%
  \BibitemOpen
  \bibfield  {author} {\bibinfo {author} {\bibfnamefont {Y.}~\bibnamefont
  {Oreg}}, \bibinfo {author} {\bibfnamefont {G.}~\bibnamefont {Refael}}, \ and\
  \bibinfo {author} {\bibfnamefont {F.}~\bibnamefont {von Oppen}},\ }\href
  {\doibase 10.1103/PhysRevLett.105.177002} {\bibfield  {journal} {\bibinfo
  {journal} {Phys. Rev. Lett.}\ }\textbf {\bibinfo {volume} {105}},\ \bibinfo
  {pages} {177002} (\bibinfo {year} {2010})}\BibitemShut {NoStop}%
\bibitem [{\citenamefont {Mourik}\ \emph {et~al.}(2012)\citenamefont {Mourik},
  \citenamefont {Zuo}, \citenamefont {Frolov}, \citenamefont {Plissard},
  \citenamefont {Bakkers},\ and\ \citenamefont {Kouwenhoven}}]{Mourik:S12}%
  \BibitemOpen
  \bibfield  {author} {\bibinfo {author} {\bibfnamefont {V.}~\bibnamefont
  {Mourik}}, \bibinfo {author} {\bibfnamefont {K.}~\bibnamefont {Zuo}},
  \bibinfo {author} {\bibfnamefont {S.}~\bibnamefont {Frolov}}, \bibinfo
  {author} {\bibfnamefont {S.}~\bibnamefont {Plissard}}, \bibinfo {author}
  {\bibfnamefont {E.}~\bibnamefont {Bakkers}}, \ and\ \bibinfo {author}
  {\bibfnamefont {L.}~\bibnamefont {Kouwenhoven}},\ }\href
  {http://www.sciencemag.org/content/early/2012/04/11/science.1222360.abstract}
  {\bibfield  {journal} {\bibinfo  {journal} {Science}\ }\textbf {\bibinfo
  {volume} {336}},\ \bibinfo {pages} {1003} (\bibinfo {year}
  {2012})}\BibitemShut {NoStop}%
\bibitem [{\citenamefont {Choy}\ \emph {et~al.}(2011)\citenamefont {Choy},
  \citenamefont {Edge}, \citenamefont {Akhmerov},\ and\ \citenamefont
  {Beenakker}}]{PhysRevB.84.195442}%
  \BibitemOpen
  \bibfield  {author} {\bibinfo {author} {\bibfnamefont {T.-P.}\ \bibnamefont
  {Choy}}, \bibinfo {author} {\bibfnamefont {J.~M.}\ \bibnamefont {Edge}},
  \bibinfo {author} {\bibfnamefont {A.~R.}\ \bibnamefont {Akhmerov}}, \ and\
  \bibinfo {author} {\bibfnamefont {C.~W.~J.}\ \bibnamefont {Beenakker}},\
  }\href {\doibase 10.1103/PhysRevB.84.195442} {\bibfield  {journal} {\bibinfo
  {journal} {Phys. Rev. B}\ }\textbf {\bibinfo {volume} {84}},\ \bibinfo
  {pages} {195442} (\bibinfo {year} {2011})}\BibitemShut {NoStop}%
\bibitem [{\citenamefont {Nadj-Perge}\ \emph {et~al.}(2014)\citenamefont
  {Nadj-Perge}, \citenamefont {Drozdov}, \citenamefont {Li}, \citenamefont
  {Chen}, \citenamefont {Jeon}, \citenamefont {Seo}, \citenamefont {MacDonald},
  \citenamefont {Bernevig},\ and\ \citenamefont {Yazdani}}]{Nadj-Perge602}%
  \BibitemOpen
  \bibfield  {author} {\bibinfo {author} {\bibfnamefont {S.}~\bibnamefont
  {Nadj-Perge}}, \bibinfo {author} {\bibfnamefont {I.~K.}\ \bibnamefont
  {Drozdov}}, \bibinfo {author} {\bibfnamefont {J.}~\bibnamefont {Li}},
  \bibinfo {author} {\bibfnamefont {H.}~\bibnamefont {Chen}}, \bibinfo {author}
  {\bibfnamefont {S.}~\bibnamefont {Jeon}}, \bibinfo {author} {\bibfnamefont
  {J.}~\bibnamefont {Seo}}, \bibinfo {author} {\bibfnamefont {A.~H.}\
  \bibnamefont {MacDonald}}, \bibinfo {author} {\bibfnamefont {B.~A.}\
  \bibnamefont {Bernevig}}, \ and\ \bibinfo {author} {\bibfnamefont
  {A.}~\bibnamefont {Yazdani}},\ }\href {\doibase 10.1126/science.1259327}
  {\bibfield  {journal} {\bibinfo  {journal} {Science}\ }\textbf {\bibinfo
  {volume} {346}},\ \bibinfo {pages} {602} (\bibinfo {year}
  {2014})}\BibitemShut {NoStop}%
\bibitem [{\citenamefont {Fu}\ and\ \citenamefont
  {Kane}(2009)}]{PhysRevB.79.161408}%
  \BibitemOpen
  \bibfield  {author} {\bibinfo {author} {\bibfnamefont {L.}~\bibnamefont
  {Fu}}\ and\ \bibinfo {author} {\bibfnamefont {C.~L.}\ \bibnamefont {Kane}},\
  }\href {\doibase 10.1103/PhysRevB.79.161408} {\bibfield  {journal} {\bibinfo
  {journal} {Phys. Rev. B}\ }\textbf {\bibinfo {volume} {79}},\ \bibinfo
  {pages} {161408} (\bibinfo {year} {2009})}\BibitemShut {NoStop}%
\bibitem [{\citenamefont {Sac\'{e}p\'{e}}\ \emph {et~al.}(2011)\citenamefont
  {Sac\'{e}p\'{e}}, \citenamefont {Oostinga}, \citenamefont {Li}, \citenamefont
  {Ubaldini}, \citenamefont {Couto}, \citenamefont {Giannini},\ and\
  \citenamefont {Morpurgo}}]{ncomms575}%
  \BibitemOpen
  \bibfield  {author} {\bibinfo {author} {\bibfnamefont {B.}~\bibnamefont
  {Sac\'{e}p\'{e}}}, \bibinfo {author} {\bibfnamefont {J.~B.}\ \bibnamefont
  {Oostinga}}, \bibinfo {author} {\bibfnamefont {J.}~\bibnamefont {Li}},
  \bibinfo {author} {\bibfnamefont {A.}~\bibnamefont {Ubaldini}}, \bibinfo
  {author} {\bibfnamefont {N.~J.}\ \bibnamefont {Couto}}, \bibinfo {author}
  {\bibfnamefont {E.}~\bibnamefont {Giannini}}, \ and\ \bibinfo {author}
  {\bibfnamefont {A.~F.}\ \bibnamefont {Morpurgo}},\ }\href {\doibase
  10.1038/ncomms1586} {\bibfield  {journal} {\bibinfo  {journal} {Nat.
  Commun.}\ }\textbf {\bibinfo {volume} {2}},\ \bibinfo {pages} {575} (\bibinfo
  {year} {2011})}\BibitemShut {NoStop}%
\bibitem [{\citenamefont {Aguado}(2017)}]{Aguadoreview17}%
  \BibitemOpen
  \bibfield  {author} {\bibinfo {author} {\bibfnamefont {R.}~\bibnamefont
  {Aguado}},\ }\href@noop {} {\bibfield  {journal} {\bibinfo  {journal} {La
  Rivista del Nuovo Cimento}\ }\textbf {\bibinfo {volume} {40}},\ \bibinfo
  {pages} {523} (\bibinfo {year} {2017})}\BibitemShut {NoStop}%
\bibitem [{\citenamefont {Lutchyn}\ \emph {et~al.}(2018)\citenamefont
  {Lutchyn}, \citenamefont {Bakkers}, \citenamefont {Kouwenhoven},
  \citenamefont {Krogstrup}, \citenamefont {Marcus},\ and\ \citenamefont
  {Oreg}}]{LutchynReview08}%
  \BibitemOpen
  \bibfield  {author} {\bibinfo {author} {\bibfnamefont {R.~M.}\ \bibnamefont
  {Lutchyn}}, \bibinfo {author} {\bibfnamefont {E.~P. A.~M.}\ \bibnamefont
  {Bakkers}}, \bibinfo {author} {\bibfnamefont {L.~P.}\ \bibnamefont
  {Kouwenhoven}}, \bibinfo {author} {\bibfnamefont {P.}~\bibnamefont
  {Krogstrup}}, \bibinfo {author} {\bibfnamefont {C.~M.}\ \bibnamefont
  {Marcus}}, \ and\ \bibinfo {author} {\bibfnamefont {Y.}~\bibnamefont
  {Oreg}},\ }\href {https://doi.org/10.1038/s41578-018-0003-1} {\bibfield
  {journal} {\bibinfo  {journal} {Nat. Rev. Mater.}\ }\textbf {\bibinfo
  {volume} {3}},\ \bibinfo {pages} {52} (\bibinfo {year} {2018})}\BibitemShut
  {NoStop}%
\bibitem [{\citenamefont {Choi}\ \emph {et~al.}(2019)\citenamefont {Choi},
  \citenamefont {Lorente}, \citenamefont {Wiebe}, \citenamefont {von Bergmann},
  \citenamefont {Otte},\ and\ \citenamefont {Heinrich}}]{magnatoms}%
  \BibitemOpen
  \bibfield  {author} {\bibinfo {author} {\bibfnamefont {D.-J.}\ \bibnamefont
  {Choi}}, \bibinfo {author} {\bibfnamefont {N.}~\bibnamefont {Lorente}},
  \bibinfo {author} {\bibfnamefont {J.}~\bibnamefont {Wiebe}}, \bibinfo
  {author} {\bibfnamefont {K.}~\bibnamefont {von Bergmann}}, \bibinfo {author}
  {\bibfnamefont {A.~F.}\ \bibnamefont {Otte}}, \ and\ \bibinfo {author}
  {\bibfnamefont {A.~J.}\ \bibnamefont {Heinrich}},\ }\href@noop {} {\bibfield
  {journal} {\bibinfo  {journal} {arXiv:1904.09941}\ } (\bibinfo {year}
  {2019})}\BibitemShut {NoStop}%
\bibitem [{\citenamefont {Zhang}\ \emph {et~al.}(2019)\citenamefont {Zhang},
  \citenamefont {Liu}, \citenamefont {Wimmer},\ and\ \citenamefont
  {Kouwenhoven}}]{zhangreview}%
  \BibitemOpen
  \bibfield  {author} {\bibinfo {author} {\bibfnamefont {H.}~\bibnamefont
  {Zhang}}, \bibinfo {author} {\bibfnamefont {D.~E.}\ \bibnamefont {Liu}},
  \bibinfo {author} {\bibfnamefont {M.}~\bibnamefont {Wimmer}}, \ and\ \bibinfo
  {author} {\bibfnamefont {L.~P.}\ \bibnamefont {Kouwenhoven}},\ }\href@noop {}
  {\bibfield  {journal} {\bibinfo  {journal} {arXiv:1905.07882}\ } (\bibinfo
  {year} {2019})}\BibitemShut {NoStop}%
\bibitem [{\citenamefont {Culcer}\ \emph {et~al.}(2019)\citenamefont {Culcer},
  \citenamefont {Keser}, \citenamefont {Li},\ and\ \citenamefont
  {Tkachov}}]{tkachov19review}%
  \BibitemOpen
  \bibfield  {author} {\bibinfo {author} {\bibfnamefont {D.}~\bibnamefont
  {Culcer}}, \bibinfo {author} {\bibfnamefont {A.~C.}\ \bibnamefont {Keser}},
  \bibinfo {author} {\bibfnamefont {Y.}~\bibnamefont {Li}}, \ and\ \bibinfo
  {author} {\bibfnamefont {G.}~\bibnamefont {Tkachov}},\ }\href@noop {}
  {\bibfield  {journal} {\bibinfo  {journal} {arXiv:1907.10058}\ } (\bibinfo
  {year} {2019})}\BibitemShut {NoStop}%
\bibitem [{\citenamefont {Lee}\ \emph {et~al.}(2017)\citenamefont {Lee},
  \citenamefont {Lutchyn},\ and\ \citenamefont {Maciejko}}]{Lee16}%
  \BibitemOpen
  \bibfield  {author} {\bibinfo {author} {\bibfnamefont {S.-P.}\ \bibnamefont
  {Lee}}, \bibinfo {author} {\bibfnamefont {R.~M.}\ \bibnamefont {Lutchyn}}, \
  and\ \bibinfo {author} {\bibfnamefont {J.}~\bibnamefont {Maciejko}},\ }\href
  {\doibase 10.1103/PhysRevB.95.184506} {\bibfield  {journal} {\bibinfo
  {journal} {Phys. Rev. B}\ }\textbf {\bibinfo {volume} {95}},\ \bibinfo
  {pages} {184506} (\bibinfo {year} {2017})}\BibitemShut {NoStop}%
\bibitem [{\citenamefont {Fu}(2010)}]{PhysRevLett.104.056402}%
  \BibitemOpen
  \bibfield  {author} {\bibinfo {author} {\bibfnamefont {L.}~\bibnamefont
  {Fu}},\ }\href {\doibase 10.1103/PhysRevLett.104.056402} {\bibfield
  {journal} {\bibinfo  {journal} {Phys. Rev. Lett.}\ }\textbf {\bibinfo
  {volume} {104}},\ \bibinfo {pages} {056402} (\bibinfo {year}
  {2010})}\BibitemShut {NoStop}%
\bibitem [{\citenamefont {Law}\ \emph {et~al.}(2009)\citenamefont {Law},
  \citenamefont {Lee},\ and\ \citenamefont {Ng}}]{PhysRevLett.103.237001}%
  \BibitemOpen
  \bibfield  {author} {\bibinfo {author} {\bibfnamefont {K.~T.}\ \bibnamefont
  {Law}}, \bibinfo {author} {\bibfnamefont {P.~A.}\ \bibnamefont {Lee}}, \ and\
  \bibinfo {author} {\bibfnamefont {T.~K.}\ \bibnamefont {Ng}},\ }\href
  {\doibase 10.1103/PhysRevLett.103.237001} {\bibfield  {journal} {\bibinfo
  {journal} {Phys. Rev. Lett.}\ }\textbf {\bibinfo {volume} {103}},\ \bibinfo
  {pages} {237001} (\bibinfo {year} {2009})}\BibitemShut {NoStop}%
\bibitem [{\citenamefont {Vijay}\ and\ \citenamefont
  {Fu}(2016)}]{PhysRevB.94.235446}%
  \BibitemOpen
  \bibfield  {author} {\bibinfo {author} {\bibfnamefont {S.}~\bibnamefont
  {Vijay}}\ and\ \bibinfo {author} {\bibfnamefont {L.}~\bibnamefont {Fu}},\
  }\href {\doibase 10.1103/PhysRevB.94.235446} {\bibfield  {journal} {\bibinfo
  {journal} {Phys. Rev. B}\ }\textbf {\bibinfo {volume} {94}},\ \bibinfo
  {pages} {235446} (\bibinfo {year} {2016})}\BibitemShut {NoStop}%
\bibitem [{\citenamefont {Hoffman}\ \emph {et~al.}(2017)\citenamefont
  {Hoffman}, \citenamefont {Chevallier}, \citenamefont {Loss},\ and\
  \citenamefont {Klinovaja}}]{PhysRevB.96.045440}%
  \BibitemOpen
  \bibfield  {author} {\bibinfo {author} {\bibfnamefont {S.}~\bibnamefont
  {Hoffman}}, \bibinfo {author} {\bibfnamefont {D.}~\bibnamefont {Chevallier}},
  \bibinfo {author} {\bibfnamefont {D.}~\bibnamefont {Loss}}, \ and\ \bibinfo
  {author} {\bibfnamefont {J.}~\bibnamefont {Klinovaja}},\ }\href {\doibase
  10.1103/PhysRevB.96.045440} {\bibfield  {journal} {\bibinfo  {journal} {Phys.
  Rev. B}\ }\textbf {\bibinfo {volume} {96}},\ \bibinfo {pages} {045440}
  (\bibinfo {year} {2017})}\BibitemShut {NoStop}%
\bibitem [{\citenamefont {Klinovaja}\ and\ \citenamefont
  {Loss}(2012)}]{PhysRevB.86.085408}%
  \BibitemOpen
  \bibfield  {author} {\bibinfo {author} {\bibfnamefont {J.}~\bibnamefont
  {Klinovaja}}\ and\ \bibinfo {author} {\bibfnamefont {D.}~\bibnamefont
  {Loss}},\ }\href {\doibase 10.1103/PhysRevB.86.085408} {\bibfield  {journal}
  {\bibinfo  {journal} {Phys. Rev. B}\ }\textbf {\bibinfo {volume} {86}},\
  \bibinfo {pages} {085408} (\bibinfo {year} {2012})}\BibitemShut {NoStop}%
\bibitem [{\citenamefont {Prada}\ \emph {et~al.}(2012)\citenamefont {Prada},
  \citenamefont {San-Jose},\ and\ \citenamefont {Aguado}}]{PhysRevB.86.180503}%
  \BibitemOpen
  \bibfield  {author} {\bibinfo {author} {\bibfnamefont {E.}~\bibnamefont
  {Prada}}, \bibinfo {author} {\bibfnamefont {P.}~\bibnamefont {San-Jose}}, \
  and\ \bibinfo {author} {\bibfnamefont {R.}~\bibnamefont {Aguado}},\ }\href
  {\doibase 10.1103/PhysRevB.86.180503} {\bibfield  {journal} {\bibinfo
  {journal} {Phys. Rev. B}\ }\textbf {\bibinfo {volume} {86}},\ \bibinfo
  {pages} {180503} (\bibinfo {year} {2012})}\BibitemShut {NoStop}%
\bibitem [{\citenamefont {Das~Sarma}\ \emph {et~al.}(2012)\citenamefont
  {Das~Sarma}, \citenamefont {Sau},\ and\ \citenamefont
  {Stanescu}}]{PhysRevB.86.220506}%
  \BibitemOpen
  \bibfield  {author} {\bibinfo {author} {\bibfnamefont {S.}~\bibnamefont
  {Das~Sarma}}, \bibinfo {author} {\bibfnamefont {J.~D.}\ \bibnamefont {Sau}},
  \ and\ \bibinfo {author} {\bibfnamefont {T.~D.}\ \bibnamefont {Stanescu}},\
  }\href {\doibase 10.1103/PhysRevB.86.220506} {\bibfield  {journal} {\bibinfo
  {journal} {Phys. Rev. B}\ }\textbf {\bibinfo {volume} {86}},\ \bibinfo
  {pages} {220506} (\bibinfo {year} {2012})}\BibitemShut {NoStop}%
\bibitem [{\citenamefont {Rainis}\ \emph {et~al.}(2013)\citenamefont {Rainis},
  \citenamefont {Trifunovic}, \citenamefont {Klinovaja},\ and\ \citenamefont
  {Loss}}]{PhysRevB.87.024515}%
  \BibitemOpen
  \bibfield  {author} {\bibinfo {author} {\bibfnamefont {D.}~\bibnamefont
  {Rainis}}, \bibinfo {author} {\bibfnamefont {L.}~\bibnamefont {Trifunovic}},
  \bibinfo {author} {\bibfnamefont {J.}~\bibnamefont {Klinovaja}}, \ and\
  \bibinfo {author} {\bibfnamefont {D.}~\bibnamefont {Loss}},\ }\href {\doibase
  10.1103/PhysRevB.87.024515} {\bibfield  {journal} {\bibinfo  {journal} {Phys.
  Rev. B}\ }\textbf {\bibinfo {volume} {87}},\ \bibinfo {pages} {024515}
  (\bibinfo {year} {2013})}\BibitemShut {NoStop}%
\bibitem [{Note1()}]{Note1}%
  \BibitemOpen
  \bibinfo {note} {MZMs always appear in pairs, one at each end of a
  topological SC wire. If the length of this wire is less than twice the
  Majorana localization length, then the Majorana wavefunctions overlap,
  leading to a finite energy splitting $\delta $. Here, we couple only one
  these MZMs to the SPW but still consider finite $\delta $.}\BibitemShut
  {Stop}%
\bibitem [{\citenamefont {Tanaka}\ \emph {et~al.}(2012)\citenamefont {Tanaka},
  \citenamefont {Sato},\ and\ \citenamefont {Nagaosa}}]{Nagaosa12}%
  \BibitemOpen
  \bibfield  {author} {\bibinfo {author} {\bibfnamefont {Y.}~\bibnamefont
  {Tanaka}}, \bibinfo {author} {\bibfnamefont {M.}~\bibnamefont {Sato}}, \ and\
  \bibinfo {author} {\bibfnamefont {N.}~\bibnamefont {Nagaosa}},\ }\href
  {\doibase http://dx.doi.org/10.1143/JPSJ.81.011013} {\bibfield  {journal}
  {\bibinfo  {journal} {J. Phys. Soc. Jpn.}\ }\textbf {\bibinfo {volume}
  {81}},\ \bibinfo {pages} {011013} (\bibinfo {year} {2012})}\BibitemShut
  {NoStop}%
\bibitem [{\citenamefont {Cayao}\ and\ \citenamefont
  {Black-Schaffer}(2017)}]{PhysRevB.96.155426}%
  \BibitemOpen
  \bibfield  {author} {\bibinfo {author} {\bibfnamefont {J.}~\bibnamefont
  {Cayao}}\ and\ \bibinfo {author} {\bibfnamefont {A.~M.}\ \bibnamefont
  {Black-Schaffer}},\ }\href {\doibase 10.1103/PhysRevB.96.155426} {\bibfield
  {journal} {\bibinfo  {journal} {Phys. Rev. B}\ }\textbf {\bibinfo {volume}
  {96}},\ \bibinfo {pages} {155426} (\bibinfo {year} {2017})}\BibitemShut
  {NoStop}%
\bibitem [{\citenamefont {Cayao}\ and\ \citenamefont
  {Black-Schaffer}(2018)}]{PhysRevB.98.075425}%
  \BibitemOpen
  \bibfield  {author} {\bibinfo {author} {\bibfnamefont {J.}~\bibnamefont
  {Cayao}}\ and\ \bibinfo {author} {\bibfnamefont {A.~M.}\ \bibnamefont
  {Black-Schaffer}},\ }\href {\doibase 10.1103/PhysRevB.98.075425} {\bibfield
  {journal} {\bibinfo  {journal} {Phys. Rev. B}\ }\textbf {\bibinfo {volume}
  {98}},\ \bibinfo {pages} {075425} (\bibinfo {year} {2018})}\BibitemShut
  {NoStop}%
\bibitem [{\citenamefont {Fleckenstein}\ \emph {et~al.}(2018)\citenamefont
  {Fleckenstein}, \citenamefont {Ziani},\ and\ \citenamefont
  {Trauzettel}}]{PhysRevB.97.134523}%
  \BibitemOpen
  \bibfield  {author} {\bibinfo {author} {\bibfnamefont {C.}~\bibnamefont
  {Fleckenstein}}, \bibinfo {author} {\bibfnamefont {N.~T.}\ \bibnamefont
  {Ziani}}, \ and\ \bibinfo {author} {\bibfnamefont {B.}~\bibnamefont
  {Trauzettel}},\ }\href {\doibase 10.1103/PhysRevB.97.134523} {\bibfield
  {journal} {\bibinfo  {journal} {Phys. Rev. B}\ }\textbf {\bibinfo {volume}
  {97}},\ \bibinfo {pages} {134523} (\bibinfo {year} {2018})}\BibitemShut
  {NoStop}%
\bibitem [{\citenamefont {Takagi}\ \emph {et~al.}(2018)\citenamefont {Takagi},
  \citenamefont {Tamura},\ and\ \citenamefont {Tanaka}}]{Tanaka19}%
  \BibitemOpen
  \bibfield  {author} {\bibinfo {author} {\bibfnamefont {D.}~\bibnamefont
  {Takagi}}, \bibinfo {author} {\bibfnamefont {S.}~\bibnamefont {Tamura}}, \
  and\ \bibinfo {author} {\bibfnamefont {Y.}~\bibnamefont {Tanaka}},\
  }\href@noop {} {\bibfield  {journal} {\bibinfo  {journal} {arXiv:1809.09324}\
  } (\bibinfo {year} {2018})}\BibitemShut {NoStop}%
\bibitem [{\citenamefont {Tamura}\ \emph {et~al.}(2019)\citenamefont {Tamura},
  \citenamefont {Hoshino},\ and\ \citenamefont {Tanaka}}]{PhysRevB.99.184512}%
  \BibitemOpen
  \bibfield  {author} {\bibinfo {author} {\bibfnamefont {S.}~\bibnamefont
  {Tamura}}, \bibinfo {author} {\bibfnamefont {S.}~\bibnamefont {Hoshino}}, \
  and\ \bibinfo {author} {\bibfnamefont {Y.}~\bibnamefont {Tanaka}},\ }\href
  {\doibase 10.1103/PhysRevB.99.184512} {\bibfield  {journal} {\bibinfo
  {journal} {Phys. Rev. B}\ }\textbf {\bibinfo {volume} {99}},\ \bibinfo
  {pages} {184512} (\bibinfo {year} {2019})}\BibitemShut {NoStop}%
\bibitem [{\citenamefont {Tsintzis}\ \emph {et~al.}(2019)\citenamefont
  {Tsintzis}, \citenamefont {Black-Schaffer},\ and\ \citenamefont
  {Cayao}}]{thanos2019}%
  \BibitemOpen
  \bibfield  {author} {\bibinfo {author} {\bibfnamefont {A.}~\bibnamefont
  {Tsintzis}}, \bibinfo {author} {\bibfnamefont {A.~M.}\ \bibnamefont
  {Black-Schaffer}}, \ and\ \bibinfo {author} {\bibfnamefont {J.}~\bibnamefont
  {Cayao}},\ }\href {\doibase 10.1103/PhysRevB.100.115433} {\bibfield
  {journal} {\bibinfo  {journal} {Phys. Rev. B}\ }\textbf {\bibinfo {volume}
  {100}},\ \bibinfo {pages} {115433} (\bibinfo {year} {2019})}\BibitemShut
  {NoStop}%
\bibitem [{\citenamefont {Bergeret}\ \emph {et~al.}(2005)\citenamefont
  {Bergeret}, \citenamefont {Volkov},\ and\ \citenamefont
  {Efetov}}]{RevModPhys.77.1321}%
  \BibitemOpen
  \bibfield  {author} {\bibinfo {author} {\bibfnamefont {F.~S.}\ \bibnamefont
  {Bergeret}}, \bibinfo {author} {\bibfnamefont {A.~F.}\ \bibnamefont
  {Volkov}}, \ and\ \bibinfo {author} {\bibfnamefont {K.~B.}\ \bibnamefont
  {Efetov}},\ }\href {\doibase 10.1103/RevModPhys.77.1321} {\bibfield
  {journal} {\bibinfo  {journal} {Rev. Mod. Phys.}\ }\textbf {\bibinfo {volume}
  {77}},\ \bibinfo {pages} {1321} (\bibinfo {year} {2005})}\BibitemShut
  {NoStop}%
\bibitem [{Note2()}]{Note2}%
  \BibitemOpen
  \bibinfo {note} {According to the sampling theorem, the shortest wavelength
  that may be reproduced by a discrete set of values a distance $b/z$ apart is
  $2b/z$, which, for $z = 10$, is equal to $1/5$ of the MZM
  distance.}\BibitemShut {Stop}%
\end{thebibliography}%
\end{document}